\documentclass[fleqn,usenatbib]{mnras}

\usepackage{fix-cm}
\usepackage{lmodern}
\usepackage{newtxtext,newtxmath}
\usepackage{colortbl}

\DeclareRobustCommand{\VAN}[3]{#2}
\let\VANthebibliography\thebibliography
\def\thebibliography{\DeclareRobustCommand{\VAN}[3]{##3}\VANthebibliography}

\usepackage{graphicx}	% Including figure files
\usepackage{amsmath}	% Advanced maths commands
\usepackage{mathtools}
\usepackage{xcolor}
\usepackage{soul}
\usepackage{tikz}
\usepackage{graphicx}
\usepackage{subcaption}
\usepackage[normalem]{ulem}
\usepackage{multirow}
\usepackage{hyperref}
\hypersetup{
    colorlinks = true,
    linkcolor = blue
    }
\usepackage[
        noabbrev,
        capitalise,
        nameinlink,]{cleveref}
\usetikzlibrary{positioning, arrows.meta, shapes.multipart}

\title[Data-driven solutions for lens detection]{Reducing false positives in strong lens detection through effective augmentation and ensemble learning}

\author[S. Rezaei et al.]{Samira Rezaei$^{1,2}$\thanks{rezaei@strw.leidenuniv.nl}, Amirmohammad Chegeni$^{1,3,4}$, Bharath Chowdhary Nagam$^{5}$, J. P. McKean$^{5,6,7}$,
\newauthor Mitra Baratchi$^{2}$, Koen Kuijken$^{1}$, and Léon V. E. Koopmans$^{5}$\\
$^{1}$Leiden Observatory, Leiden University,  2333 CC Leiden, the Netherlands.\\
$^{2}$Leiden Institute of Advanced Computer Science (LIACS), Leiden University, 2333 CC Leiden, the Netherlands.\\
$^{3}$Dipartimento di Fisica e Astronomia "G. Galilei", Università di
Padova, Via Marzolo 8, 35131 Padova, Italy\\
$^{4}$INFN-Padova, Via Marzolo 8, 35131 Padova, Italy\\
$^{5}$Kapteyn Astronomical Institute, University of Groningen, Postbus 800, NL9700 AV Groningen, the Netherlands\\
$^{6}$South African Radio Astronomy Observatory (SARAO), P.O. Box 443, Krugersdorp 1740, South Africa\\
$^{7}$Department of Physics, University of Pretoria, Lynnwood Road, Hatfield, Pretoria, 0083, South Africa
}

\date{Accepted XXX. Received YYY; in original form 2024 September 2}

\begin{document}
\label{firstpage}
\pagerange{\pageref{firstpage}--\pageref{lastpage}}
\maketitle

\begin{abstract}
This research studies the impact of high-quality training datasets on the performance of Convolutional Neural Networks (CNNs) in detecting strong gravitational lenses. We stress the importance of data diversity and representativeness, demonstrating how variations in sample populations influence CNN performance. In addition to the quality of training data, our results highlight the effectiveness of various techniques, such as data augmentation and ensemble learning, in reducing false positives while maintaining model completeness at an acceptable level. This enhances the robustness of gravitational lens detection models and advancing capabilities in this field. Our experiments, employing variations of DenseNet and EfficientNet, achieved a best false positive rate (FP rate) of $10^{-4}$, while successfully identifying over 88 per cent of genuine gravitational lenses in the test dataset. This represents an 11-fold reduction in the FP rate compared to the original training dataset. Notably, this substantial enhancement in the FP rate is accompanied by only a 2.3 per cent decrease in the number of true positive samples. Validated on the KiDS dataset, our findings offer insights applicable to ongoing missions, like Euclid.
\end{abstract}


\section{Introduction}

Strong gravitational lensing occurs when light from a distant galaxy gets bent by the curvature of space-time that is caused by another galaxy along our line of sight. This phenomenon significantly impacts our understanding of the Universe. Strong gravitational lensing, as predicted by the theory of general relativity, occurs when the foreground galaxy acts as a lens, creating multiple magnified and distorted images of the background galaxy (see \citealt{Treu2010} for a review). Strong lensing serves as a unique tool for testing models for galaxy formation \citep{SLandGF} and cosmology \citep{cao2015cosmology}. Measuring the mass components of early-type galaxies, constraining their stellar initial mass function, and determining their inner mass density profiles \citep{Treu2006,Bolton2008,Auger2009,Auger2010,Spiniello2012,Spiniello2014,Wucknitz2004,Koopmans2006,Spingola2018}, probing the nature of dark matter through detailed modelling of the surface brightness distribution of lensed images \citep{Vegetti2012,Vegetti2014,Ritondale2019,Hsueh2020,Gilman2020}, testing models for the expansion of the Universe and dark energy \citep{Suyu2010,Suyu2013,Bonvin2017,Wong2020}, and measuring the Hubble constant \citep{Birrer2021} are some applications of studying this phenomenon.

Despite the profound insights gained by strong lensing, identifying gravitational lenses remains a challenge, as these events are rare, with the probability of one gravitational lens being found in about a thousand observed galaxies \citep{Chae2002,Wardlow2013,Amante2020}. Visual inspection of the extensive parent population is both time-consuming and susceptible to incompleteness \citep{Jackson2008,Marshall2016,More2016}. Consequently, the discovery of gravitational lenses often relies on the application of selection criteria in catalogue space, considering factors such as optical colour, radio spectral index, total flux density, and the morphology of candidate lensed images (see \citealt{Spingola2019} for an illustrative example). Despite the use of these criteria, some degree of visual inspection remains necessary to validate potential lens candidates. As we anticipate parent samples to surpass $10^7$ galaxies in size, driven by the data volumes from wide-field surveys conducted by the Vera C. Rubin Observatory \citep{rubin}, the Nancy Grace Roman Space Telescope \citep{roman}, and Euclid \citep{euclid}, the imperative for sophisticated automated search techniques becomes evident. Developing such techniques is crucial to efficiently navigate and analyze datasets of this scale. The literature has seen the emergence of several intelligent approaches tailored to diverse imaging surveys, reflecting the need for innovative and effective solutions in the face of increasingly massive datasets. \citet{Metcalf2019} conducted a lens finding challenge focused on optical/infrared datasets, revealing that automated approaches, including machine learning algorithms, outperformed traditional methods.

Convolutional Neural Networks (CNNs), a subset of machine learning models, have emerged as significant tools for cosmological and astronomical applications. These models excel at identifying patterns in complex datasets, making them highly effective for a range of scientific tasks. CNN methods have been applied in the study of exoplanet detection, where they enhance the identification and analysis of potential exoplanets from large datasets \citep{cnnast1, cnnast2}. In the field of radio astronomy, CNNs have been instrumental in classifying radio galaxies by analyzing their morphological features \citep{cnnast3, cnnast4}. They have also been used in the detection and analysis of gravitational waves, helping to filter noise and improve signal detection from data collected by observatories such as LIGO and Virgo \citep{cnnast5, cnnast6}. Furthermore, CNNs have contributed significantly to distinguishing alternative dark energy \citep{chegeni2024clusternets} and dark matter \citep{cnnast7} scenarios compared to the standard model of cosmology. In the study of the Cosmic Microwave Background (CMB) maps, CNNs have been employed to fill the masked regions of the CMB \citep{Sadr:2020rje}, extract cosmological parameters and identify subtle features in the CMB data \citep{Sadr:2020rje, cnnast8}. These diverse applications underscore the versatility and power of CNNs in enhancing our understanding of the Universe. A comprehensive review of CNN applications across various astronomical problems was presented by \citet{REZAEI2025100921}.

Numerous studies have investigated the application of CNNs for identifying strong gravitational lenses in vast datasets, automating a process that traditionally required extensive manual effort. For instance, \citet{Petrillo2019} and \citet{Rojas2021} showcased the efficiency of CNNs in detecting gravitational lenses in the Kilo Degree Survey (KiDS) data, while \citet{Nagam2023} recently proposed a new pipeline-ensemble model using Densely Connected Neural Networks (DenseNets) to reduce false positives in the identification of strong gravitational lenses, making it more suitable for large-scale astronomical surveys like Euclid. An additional example of using ensemble techniques for identifying gravitationally lensed quasars was presented by \citet{Andika2023}, where an ensemble averaging approach combines state-of-the-art convolutional and transformer-based neural networks applied to multi-band images.

Utilizing a CNN classifier on both single-band and multi-band KiDS images, \citet{Li2021} identified a sample of 97 new high-quality strong lensing candidates. This effort contributes to a total of 268 high-quality candidates from KiDS, optimizing classifier efficiency for future large-scale surveys. In addition, \citet{Rezaei2022} developed and validated CNNs for detecting galaxy-scale gravitational lenses in interferometric data from the International LOFAR Telescope (ILT), aiming for predicting a pure selection of lens candidates. The effectiveness of CNNs in these diverse settings highlights their robustness and adaptability, making them indispensable tools for the automated detection of strong gravitational lenses across various astronomical surveys.

Current research on CNNs has largely focused on refining model architectures and optimizing hyperparameters to improve performance metrics, as highlighted in recent reviews (e.g., \citealt{2018arXiv180708169R, menghani2023efficient}). While these technical improvements are essential, the quality of training data is equally crucial for CNN effectiveness. High-quality, diverse, and representative datasets significantly boost a model’s generalization ability and robustness. For instance, \citet{Canameras2023}
demonstrates that tailoring the training dataset in both the lensed and non-lensed samples can lead to improved results. That study shows how CNNs perform best when the dataset includes mock lenses closely resembling real, detectable systems with multiple lensed images that are bright and deblended. Such samples help the model to clearly differentiate between lensed and non-lensed objects. Likewise, including a high proportion of non-lensed contaminants (images resembling lenses without strong lensing features), improves the model's performance by learning to filter out such non-lensed objects.

By exploring the relationship between training data quality and the performance of a CNN, with a particular architecture performance, we aim to uncover insights that go beyond conventional optimization strategies. Specifically, we seek to understand how variations in training data characteristics, such as data distribution, labeling accuracy, and dataset size, can influence a CNN's ability to learn and generalize from the provided data. Additionally, we investigate whether certain architectures exhibit greater resilience to variations in training data.

The structure of this paper is organized as follows. Section~\ref{training_data} outlines the methodology for generating the training, validation, and test datasets, focusing on the creation of two primary classes: lensed and non-lensed data. In Section~\ref{method}, we delve into the exploration of different training strategies and their respective definitions. Moreover, we describe the CNN architectures utilized in this study. We conduct an analysis of several CNN architectures to demonstrate that the dependence on training data is not specific to any particular CNN architecture. Section~\ref{results} presents a comparative analysis of the results obtained from each training strategy, considering both the training data and the CNN architecture employed. Lastly, in Section~\ref{conclusion}, we engage in a thorough discussion of the results obtained in this study, providing insights and reflections on the methodology employed and suggesting potential avenues for future research and development.

\section{Training and Test Dataset}\label{training_data}

\begin{figure*}
    \centering
    \begin{subfigure}{0.16\textwidth}
        \centering
        \includegraphics[width=\linewidth]{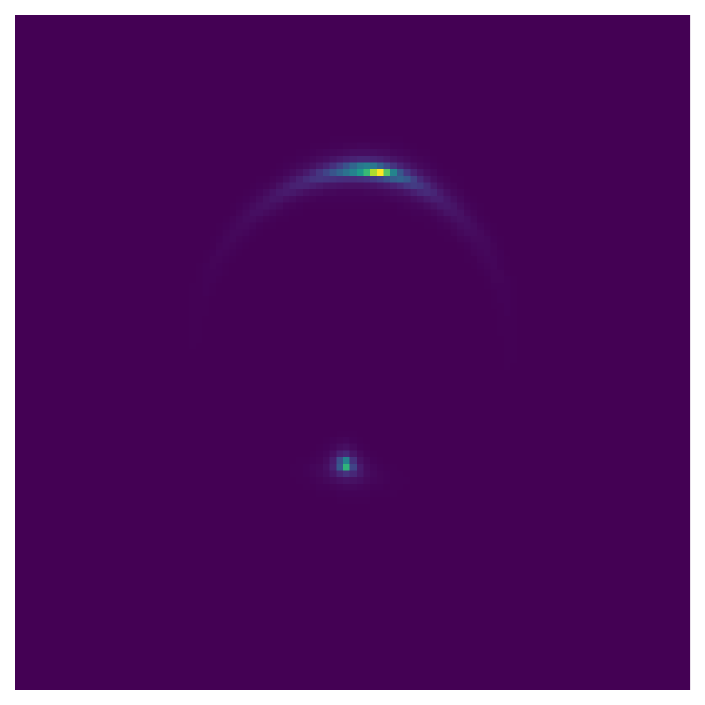}
    \end{subfigure}%
    \hfill
    \begin{subfigure}{0.16\textwidth}
        \centering
        \includegraphics[width=\linewidth]{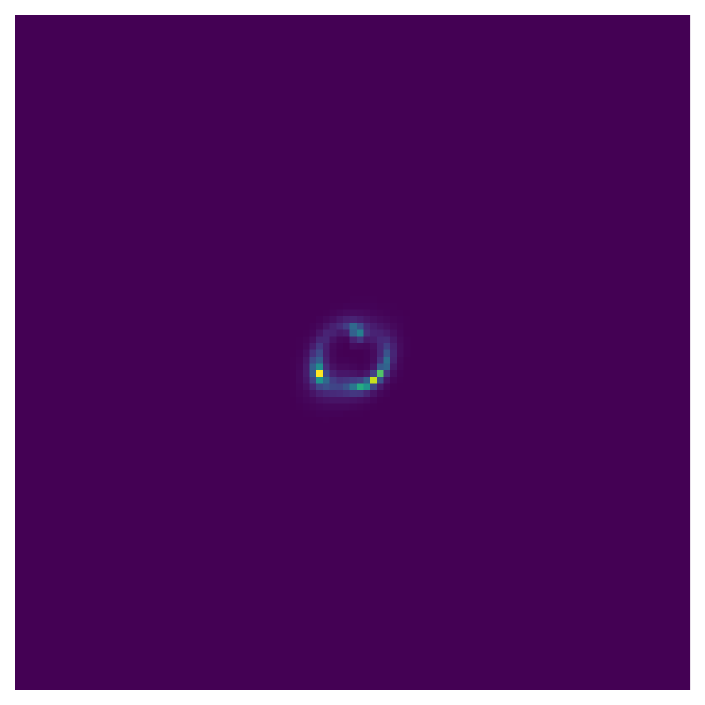}
    \end{subfigure}%
    \hfill
    \begin{subfigure}{0.16\textwidth}
        \centering
        \includegraphics[width=\linewidth]{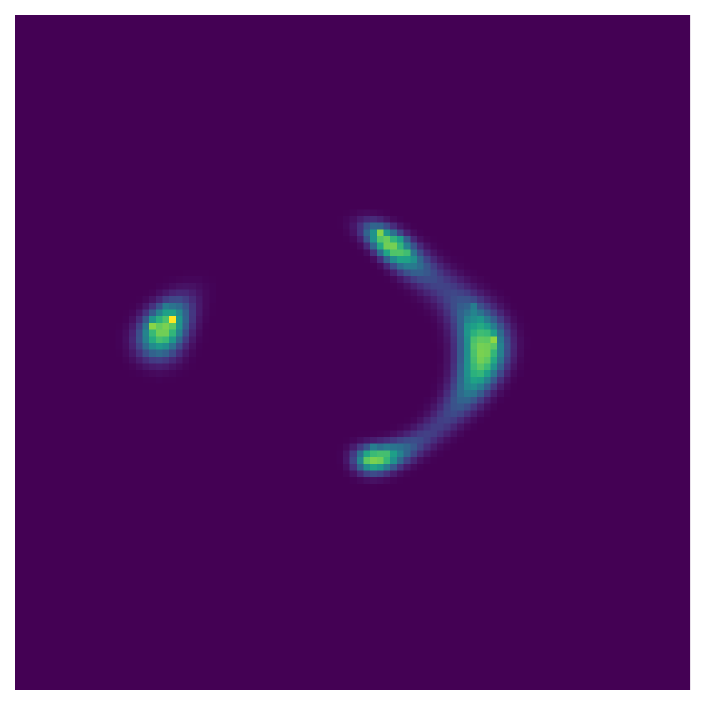}
    \end{subfigure}%
    \hfill
    \begin{subfigure}{0.16\textwidth}
        \centering
        \includegraphics[width=\linewidth]{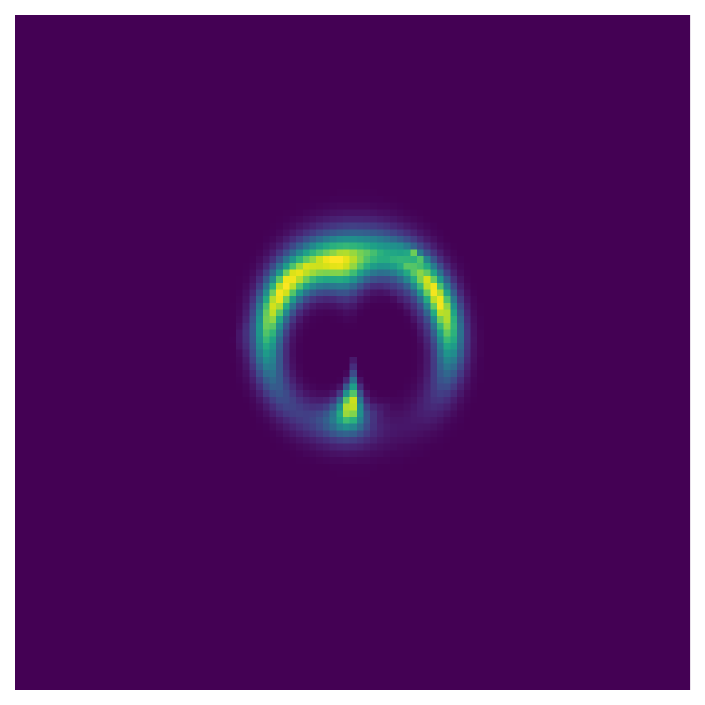}
    \end{subfigure}%
    \hfill
    \begin{subfigure}{0.16\textwidth}
        \centering
        \includegraphics[width=\linewidth]{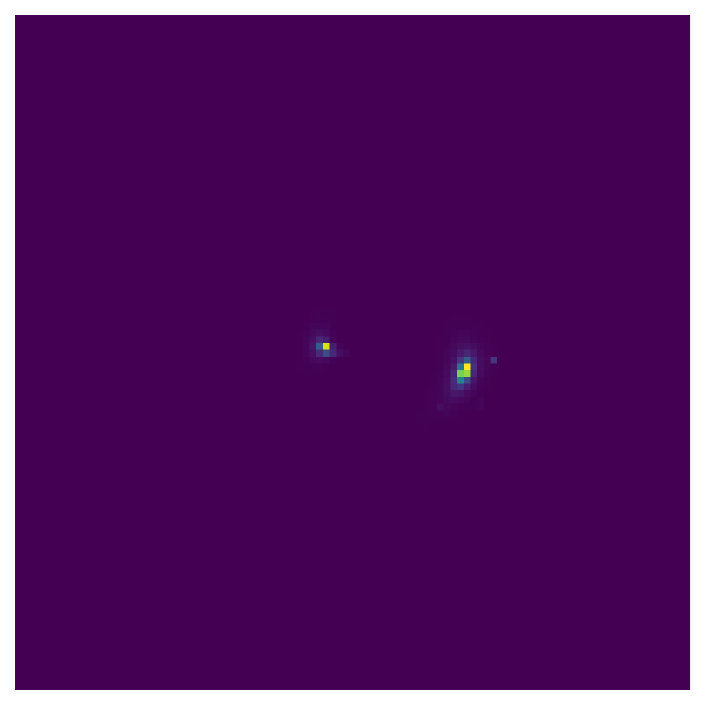}
    \end{subfigure}%
    \hfill
    \begin{subfigure}{0.16\textwidth}
        \centering
        \includegraphics[width=\linewidth]{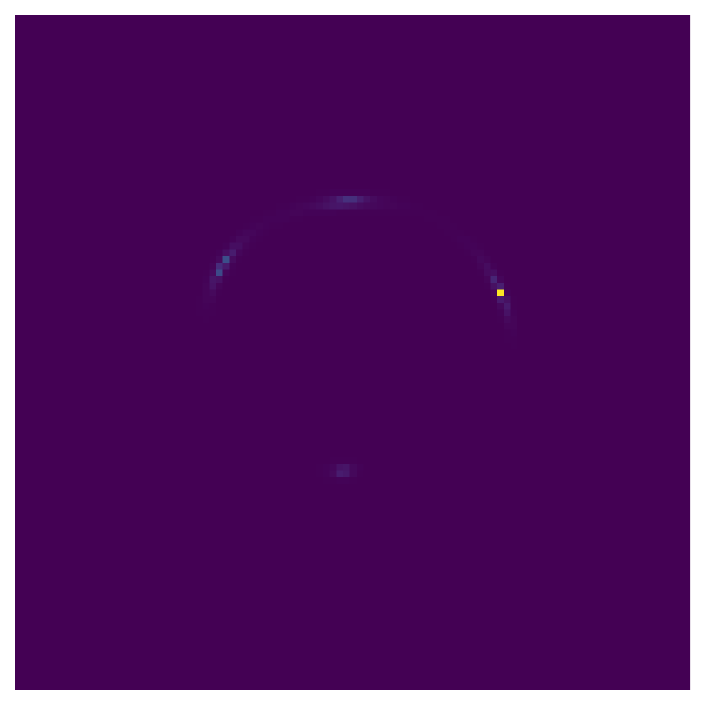}
    \end{subfigure}

    %\vspace{1em} % add some vertical space between rows

    \begin{subfigure}{0.16\textwidth}
        \centering
        \includegraphics[width=\linewidth]{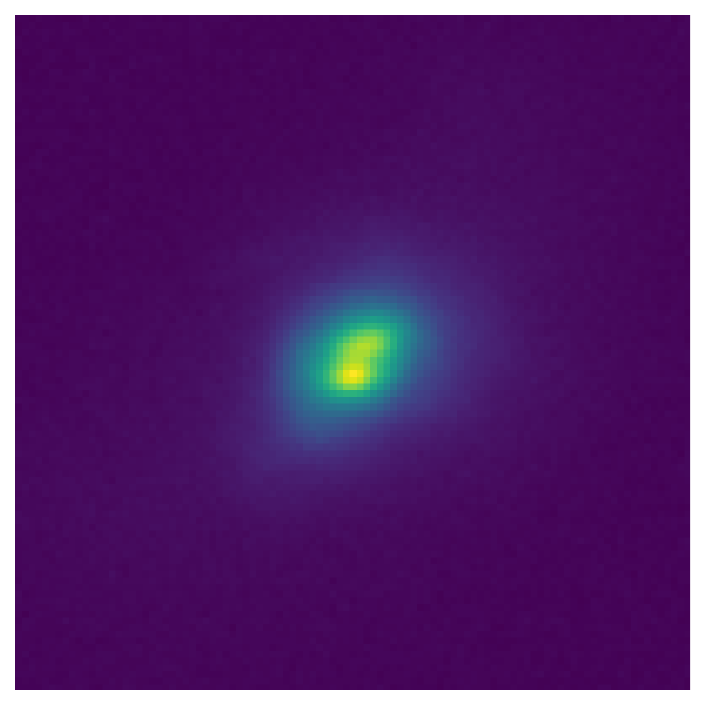}
    \end{subfigure}%
    \hfill
    \begin{subfigure}{0.16\textwidth}
        \centering
        \includegraphics[width=\linewidth]{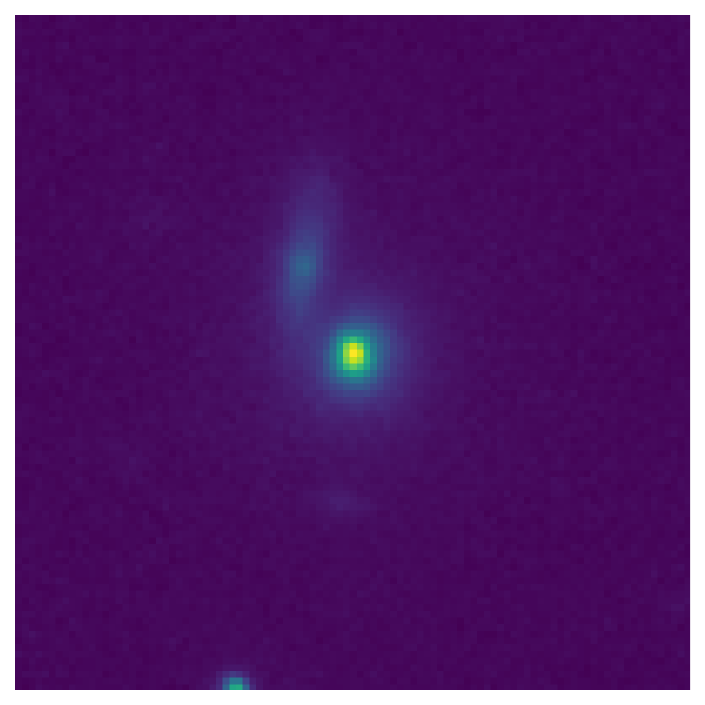}
    \end{subfigure}%
    \hfill
    \begin{subfigure}{0.16\textwidth}
        \centering
        \includegraphics[width=\linewidth]{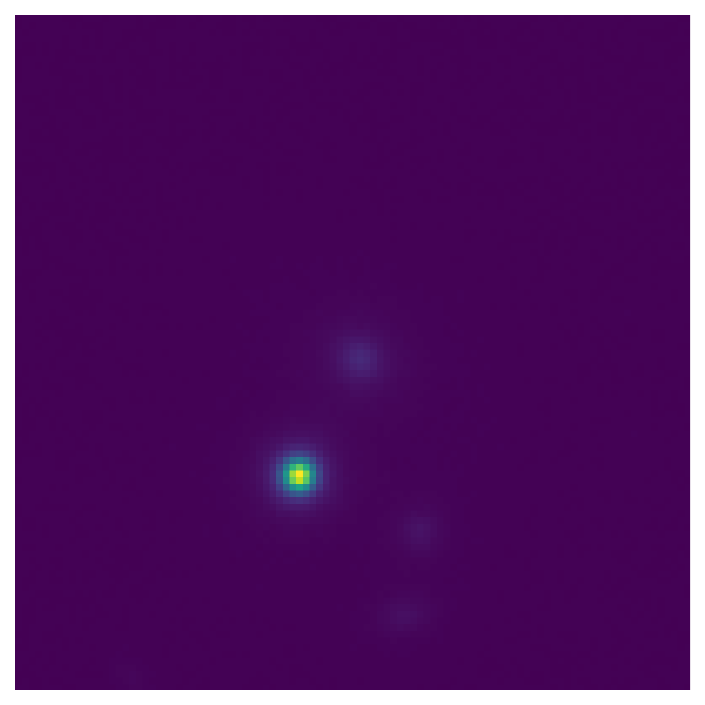}
    \end{subfigure}%
    \hfill
    \begin{subfigure}{0.16\textwidth}
        \centering
        \includegraphics[width=\linewidth]{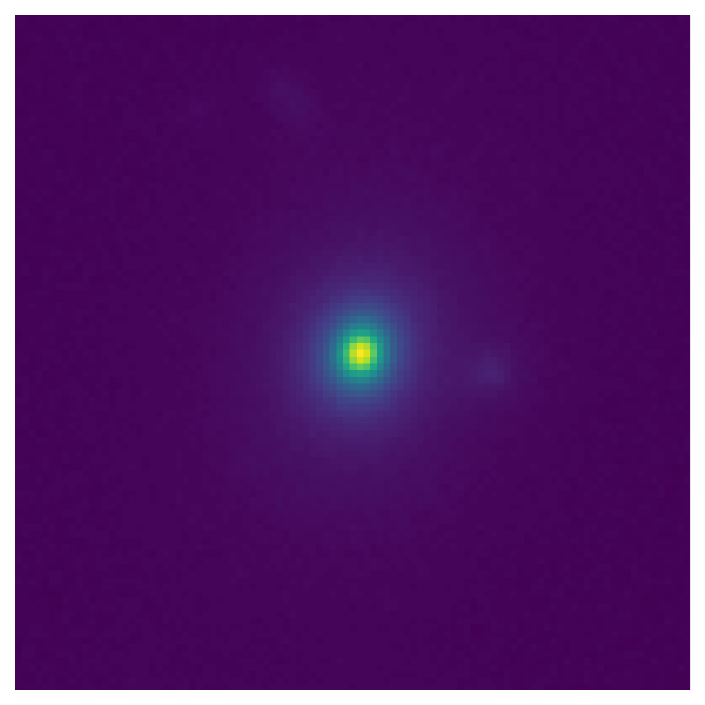}
    \end{subfigure}%
    \hfill
    \begin{subfigure}{0.16\textwidth}
        \centering
        \includegraphics[width=\linewidth]{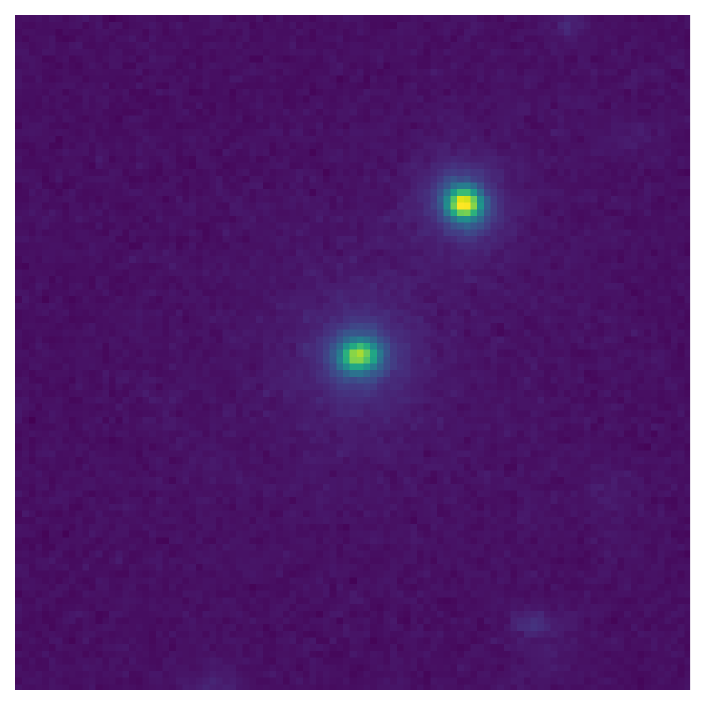}
    \end{subfigure}%
    \hfill
    \begin{subfigure}{0.16\textwidth}
        \centering
        \includegraphics[width=\linewidth]{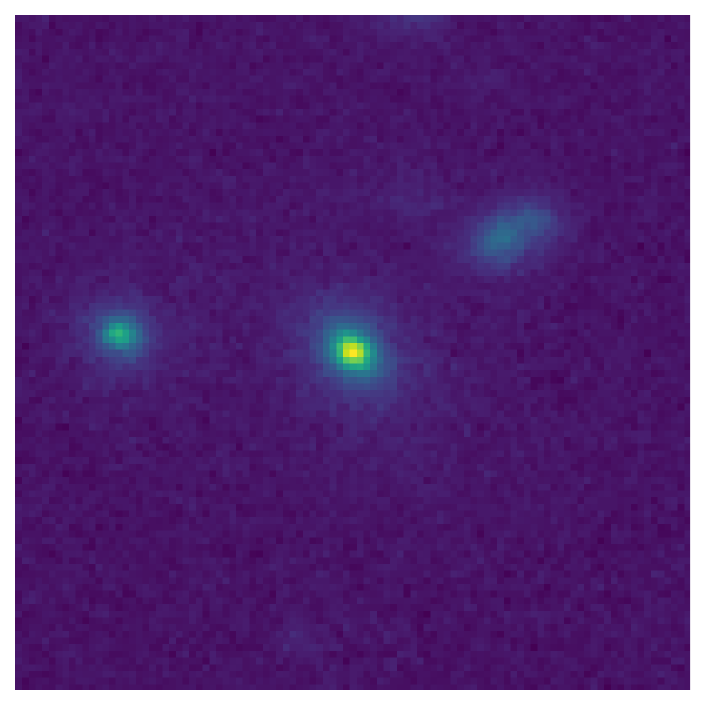}
    \end{subfigure}

    %\vspace{1em} % add some vertical space between rows

    \begin{subfigure}{0.16\textwidth}
        \centering
        \includegraphics[width=\linewidth]{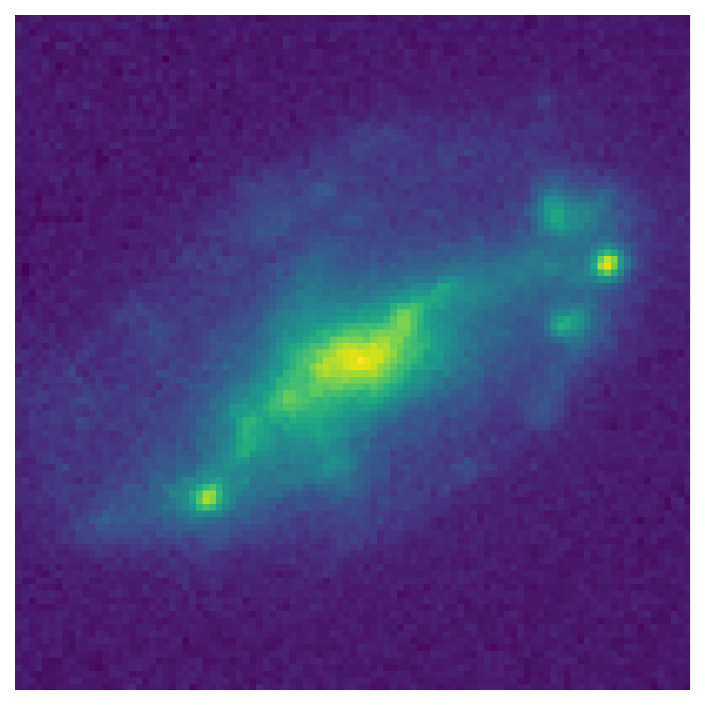}
    \end{subfigure}%
    \hfill
    \begin{subfigure}{0.16\textwidth}
        \centering
        \includegraphics[width=\linewidth]{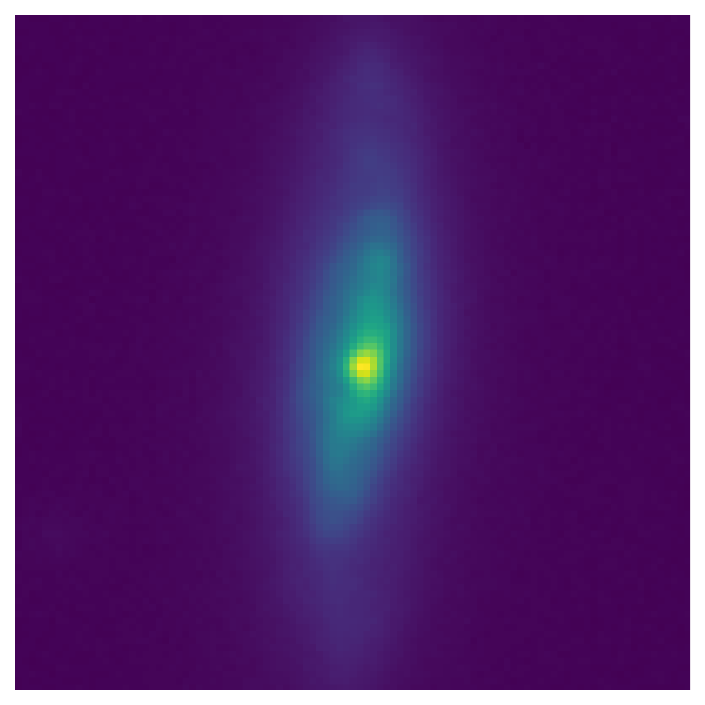}
    \end{subfigure}%
    \hfill
    \begin{subfigure}{0.16\textwidth}
        \centering
        \includegraphics[width=\linewidth]{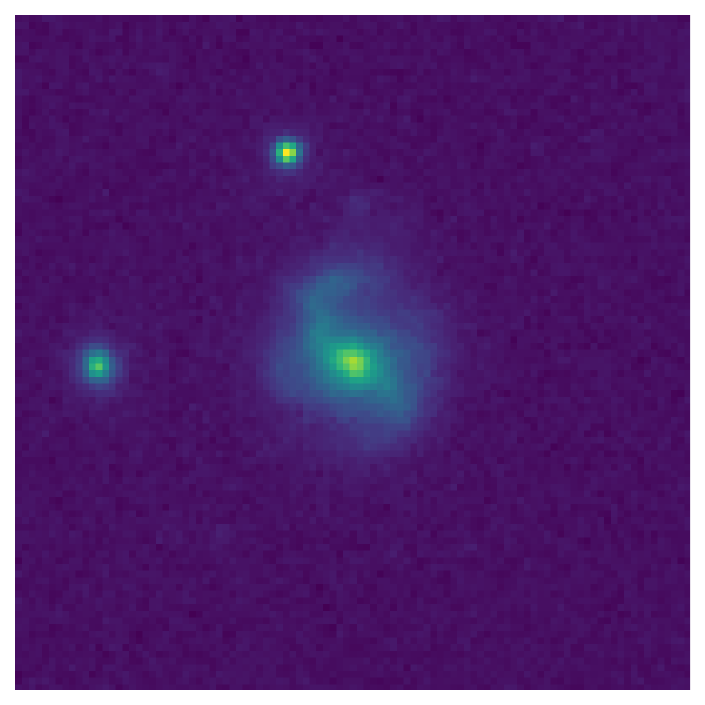}
    \end{subfigure}%
    \hfill
    \begin{subfigure}{0.16\textwidth}
        \centering
        \includegraphics[width=\linewidth]{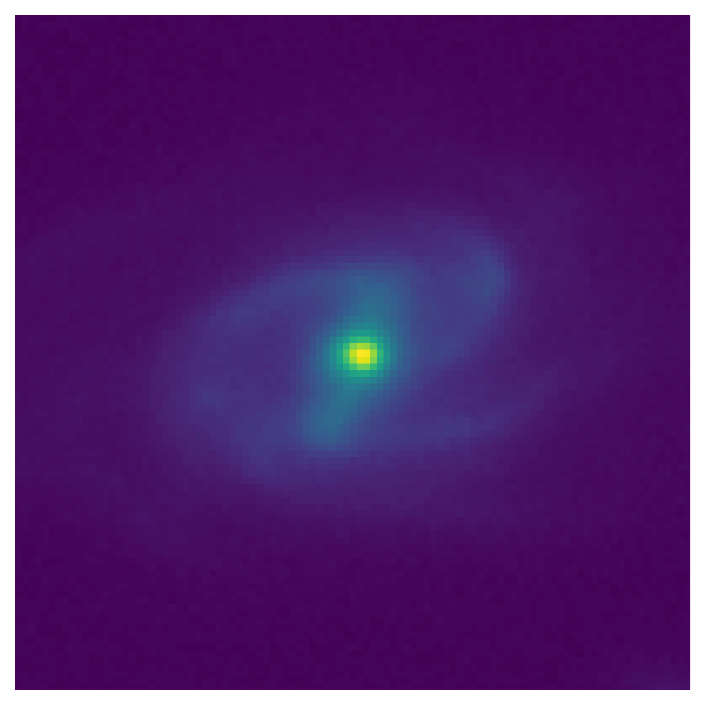}
    \end{subfigure}%
    \hfill
    \begin{subfigure}{0.16\textwidth}
        \centering
        \includegraphics[width=\linewidth]{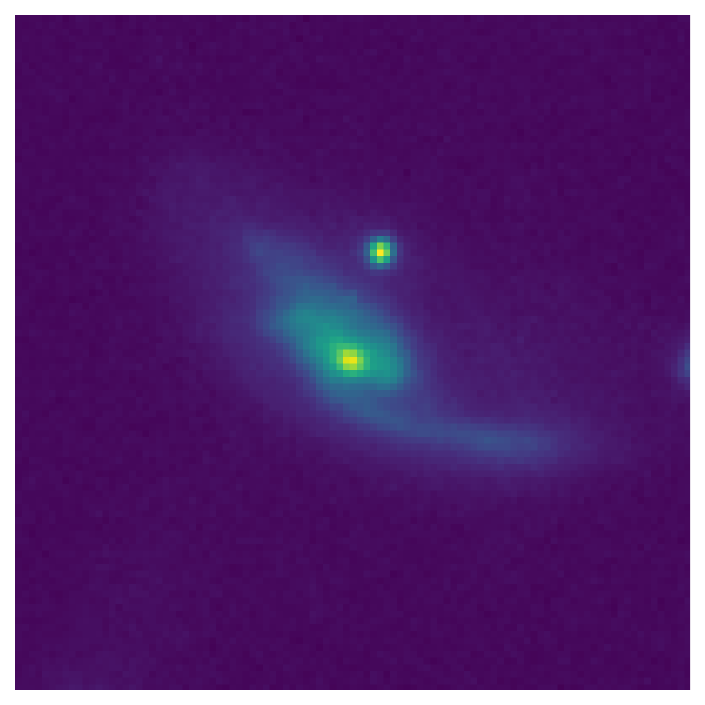}
    \end{subfigure}%
    \hfill
    \begin{subfigure}{0.16\textwidth}
        \centering
        \includegraphics[width=\linewidth]{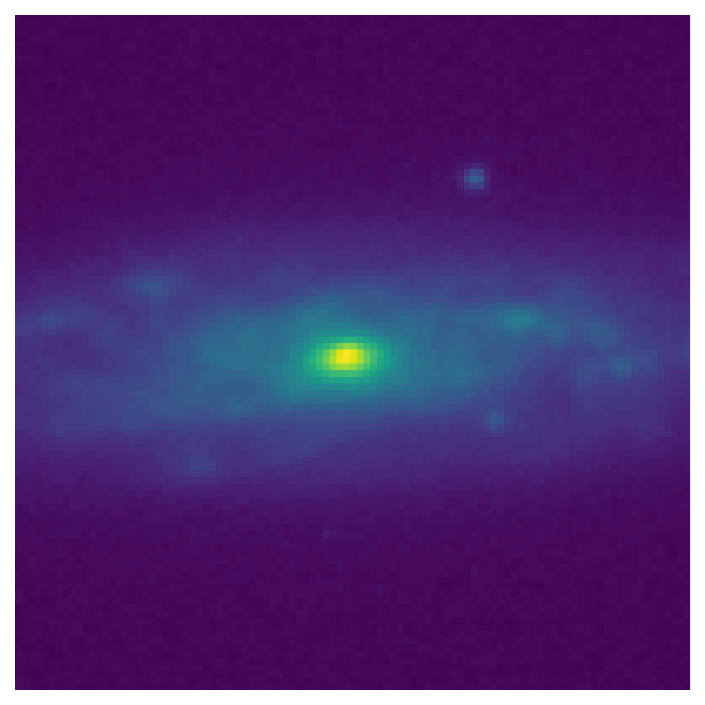}
    \end{subfigure}

    \caption{These images illustrate the dataset's diversity, presenting lensed phenomena and non-lensed galaxies. The top row showcases various lensed samples, offering insights into their morphology and configurations. The middle panel displays a selection of LRGs, serving as both foreground and non-lensed instances, while the bottom panel includes spiral galaxies. More details about the properties of these galaxies are available in Section~\ref{training_data} and Table~\ref{tab:sim_params}. Each image has a size of $101 \times 101$ pixels, which corresponds to an area of $20 \times 20$~arcsec.}
    \label{fig:lens_nonlens_samples}
\end{figure*}

In order to assemble the training, validation, and test datasets for this research, it is necessary to establish two main categories of data: "lensed" and "non-lensed" samples. A lens system is created by pairing an elliptical galaxy, specifically a Luminous Red Galaxy (LRG) from the actual KiDS data release 4 (DR4; \citealt{Kuijken2019}), denoted as the "lensing galaxy" or "foreground galaxy", with a simulated lens configuration (mock lens). The same LRGs can also be used in the non-lensed class together with a selection of spiral galaxies and contaminants collected from the KiDS catalog. A visual representation of both classes is provided in Fig.~\ref{fig:lens_nonlens_samples}. The top row exhibits a diverse array of lensed samples, showcasing various morphologies and configurations, offering a representative glimpse into the spectrum of lensed phenomena within our dataset. The middle panel displays samples of LRGs, serving as both foreground lensing galaxies and non-lensed instances. Furthermore, the non-lensed category encompasses not only LRGs, but also spiral galaxies depicted in the bottom panel of Fig.~\ref{fig:lens_nonlens_samples}.

The lensed and non-lensed samples within our dataset display a wide range of surface brightness distributions. To ensure consistency in pixel value ranges across both classes of data, we implement a \textit{MinMax} normalization. The objective of this normalization is to standardize the model by aligning the distribution of inputs. Although this normalization method results in scaling the absolute loss amplitude in training, it does not affect our capability to identify lens candidates. This is because our analysis focuses on the relative surface brightness of the lensed images and non-lensed source emission within each simulated sample. Mathematically, this normalization can be expressed as,
\begin{equation}
\centering
    x_{\rm normalized}= \frac{x-\min(x_d)}{\max(x_d)-\min(x_d)},
\label{eq:minmax}
\end{equation}
where \(x\) represents the value of a specific pixel, and \(x_d\) encompasses all pixels in an image. Through this process, all pixel values within a given image are mapped to the range of \((0, 1)\). A square-root stretch is applied to this normalized image to enhance features with lower surface brightness. 

In the following subsections, we provide detailed information on how both lensed and non-lensed classes of data are generated. 

\subsection{Non-lensed samples} \label{data_non-lens}

The task of distinguishing genuine gravitational lenses from various celestial objects in surveys like KiDS is challenging due to factors such as the diverse range of objects observed and variations in colour. Objects spanning different morphologies, including galaxies and artifacts, can potentially mimic lensing features, while limitations in survey resolution and depth may obscure faint lensed signals. To effectively train CNNs for lens detection, it is crucial to compile a comprehensive dataset that encompasses a wide variety of objects with diverse characteristics. By exposing the CNNs to this diverse dataset during training, they can learn to distinguish between genuine lensing events and contaminants, thereby improving the accuracy and reliability of lens detection methods in astronomical surveys. 

Our approach in selecting non-lensed samples aligns with previous studies, such as by \citet{Nagam2023} and \citet{Petrillo2019}. It comprises three distinct classes of data: a) 3000 LRGs with an $r$ magnitude below 21; b) 2000 sources that were previously misidentified as mock lenses in earlier tests conducted by \citet{Petrillo2017}; and c) 1000 galaxies visually classified as spiral galaxies through the GalaxyZoo project \citep{2014MNRAS.438.2882M, 2013MNRAS.435.2835W}. This diverse selection introduces a wide range of objects with different characteristics, enhancing the robustness of our training dataset for the CNNs.

We use the same backbone of data as in previous studies \citep{Petrillo2019, Nagam2023}, but we make changes to improve the CNN performance. In order to understand the data, we need to investigate deeper into the properties of the training dataset. For those 3000 LRGs (see above), we calculate the S{\'e}rsic profile \citep{sersic1968atlas}, which is a mathematical function used to describe the distribution of light in galaxies. The parameters of this profile provide insights into their structural properties such as size, magnitude/flux and morphology. By fitting the observed intensity profiles of LRGs, cropped to $20 \times 20$ pixel cutouts, with the S{\'e}rsic function, we can extract parameters such as the effective radius and S{\'e}rsic index. We use these cutouts to eliminate the effects of contamination in our S{\'e}rsic profile calculations. Our aim is to shed light on the nature and distribution of galaxies in our sample. Moreover, we introduced an additional parameter referred to as \textit{galaxy complexity}, which is defined as,
\begin{equation}
\centering
    {\rm Complexity} = \frac{S_I}{S_P},
\label{eq:compactness}
\end{equation}
where $S_I$ denotes the galaxy's integrated surface brightness, and $S_P$ represents the galaxy's peak surface brightness. This metric provides insights into the extent of an elliptical galaxy, helping us assess its size. In our experiments, we considered this property to enhance our understanding of the lensing galaxy's dimensions, a critical factor when selecting a suitable mock lensed system to be added to a lensing galaxy.

Fig.~\ref{fig:lensingGalaxyInfo} shows the distribution of the effective radius (top) and compactness (bottom) for the LRG samples to be considered as both non-lensed samples and foreground lensing galaxies. Notably, the selection of foreground lensing galaxies exhibits a significant bias toward effective radii in the range of 0.65 to 0.85 arcsec and complexity between 70 to 80. This observation proves invaluable, particularly in understanding the training data and its impact on the performance of trained CNNs. Later in Sections~\ref{applied1} and \ref{Applied2} we show how the properties of training data affect the completeness and purity of the generated lens candidates. 

\subsection{Lensed samples}

Lensed samples are created by combining simulated gravitational arcs, rings, quads, and doubles and a selection of foreground elliptical galaxies. These elliptical galaxies, which also act as the foreground lensing galaxies, are collected from the KiDS DR4, and therefore, provide a varied and representative collection of instances for both training and evaluation purposes. The process entails producing artificial distortions, such as arcs, rings, quads, and doubles, around selected foreground galaxies to replicate the appearance of gravitational lensing. The goal of these simulated lens configurations is to mimic the gravitational lensing characteristics observed in actual astronomical data, thereby providing a comprehensive and diverse dataset for training and evaluation. In the following, we explain how the "lensing galaxies" and the "mock lenses" are selected for the purpose of this work. 

\subsubsection{Foreground Galaxies} \label{lensing galaxy}

Similar to previous work, such as by \citet{Petrillo2017, Petrillo2019b} and \citet{Nagam2023}, we focus on low-redshift ($z \leq 0.4$), massive early-type galaxies (i.e. LRGs), which have been established as the predominant contributors to the lensing galaxy population (e.g. \citealt{Moller2007, Oguri2006}). The training dataset comprises 4,411 unique LRGs, with an additional 511 samples allocated for validation and another 511 for samples for testing. Further details on the selection process of LRGs to be considered as foreground galaxies is presented by  \citet{Nagam2023}.

In contrast to previous studies (e.g. \citealt{Nagam2023, Li2021, Petrillo2019}), which randomly select a mock lensed source and a foreground galaxy to form a lens system, we consider the properties of both components. This approach enables us to generate realistic gravitational lens samples, but to also provide our deep learning model with clear samples that minimize the risk of misinterpretation, which should improve the accuracy of our analysis. This would also enhance the model's ability to generalize to unseen data. More details on how we have adjusted the selection of LRGs to be used in combination with a mock lens is provided in Section~\ref{Applied2}.

\begin{figure}
    \centering

    \begin{minipage}{0.45\textwidth}
        \centering
        \includegraphics[width=\linewidth]{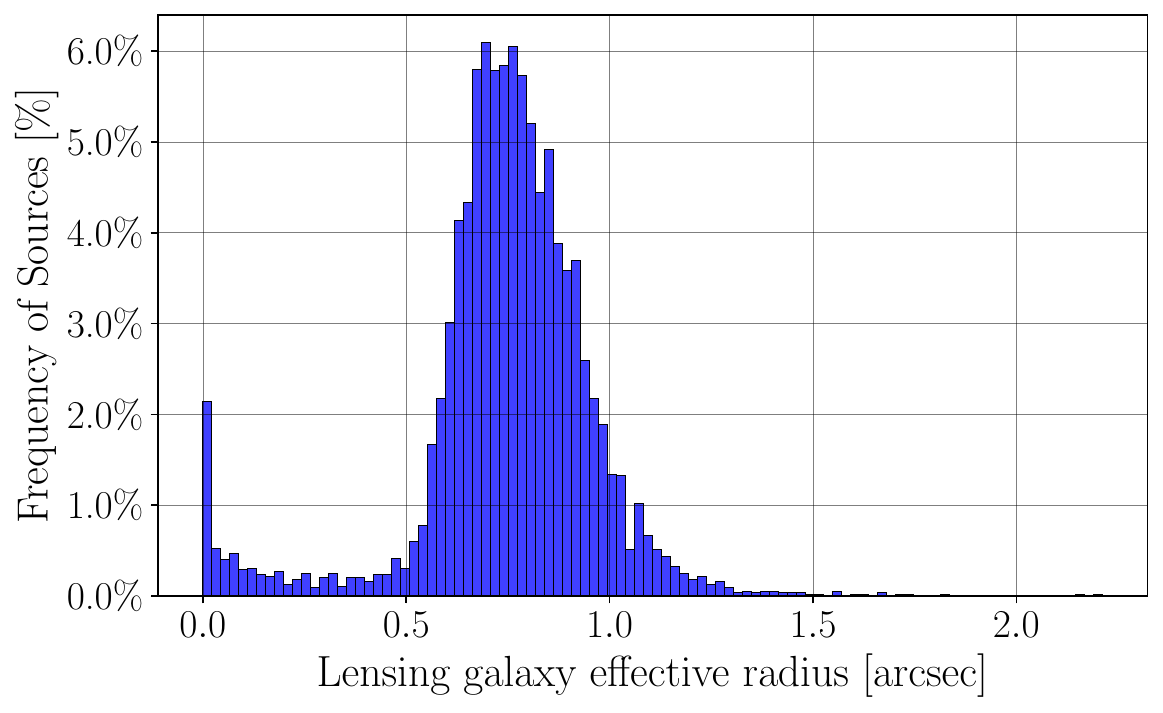}
    \end{minipage}%
    \hfill
    \vspace{0.25cm}
    \begin{minipage}{0.45\textwidth}
        \centering
        \includegraphics[width=\linewidth]{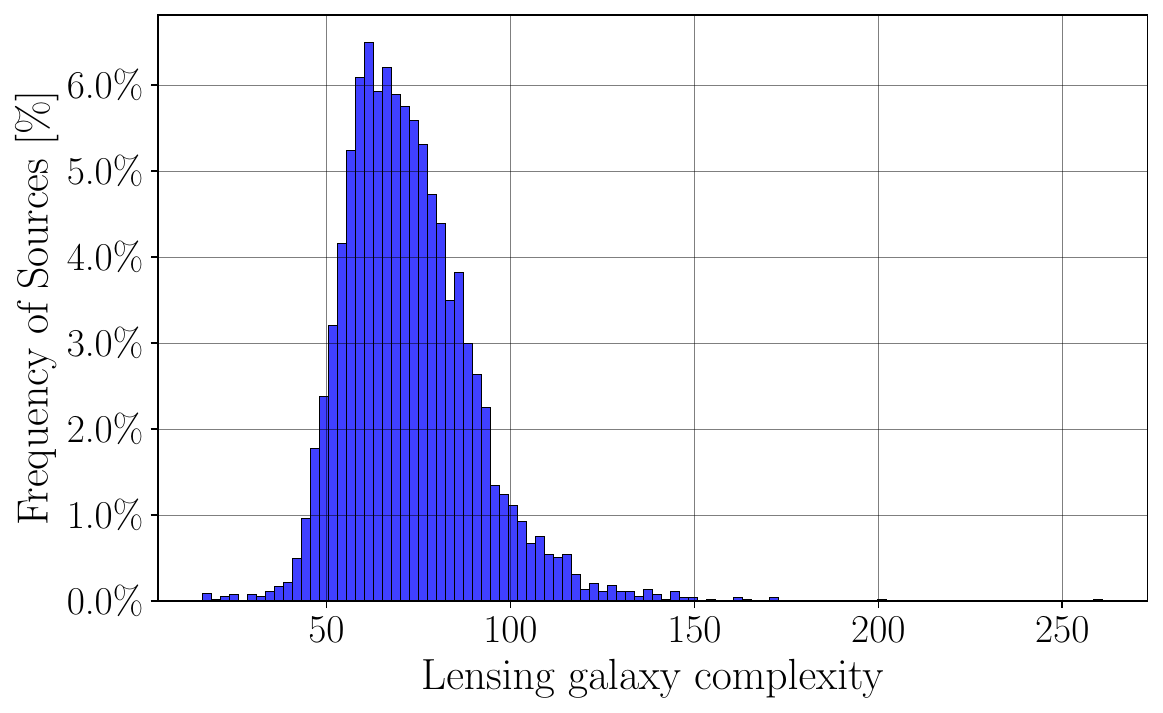}
        %\caption{Caption for Plot 2}

    \end{minipage}%

    \caption{The distribution of the lensing galaxy model parameters used for the training dataset; lensing galaxy effective radius (top) and the lensing galaxy complexity (bottom).} 
    \label{fig:lensingGalaxyInfo}
\end{figure}

\subsubsection{Mock Lenses} \label{mock_lens}
         
Developing a robust training dataset for strong gravitational lensing detection poses unique challenges, demanding representation across a diverse set of lensing configurations, while maintaining the spatial sampling of KiDS ($\sim0.2$~arcsec~pixel$^{-1}$). The selection of lens and source parameters is detailed in Table~\ref{tab:sim_params}.

\begin{figure*}
    \centering

    \begin{minipage}{0.33\textwidth}
        \centering
        \includegraphics[width=\linewidth]{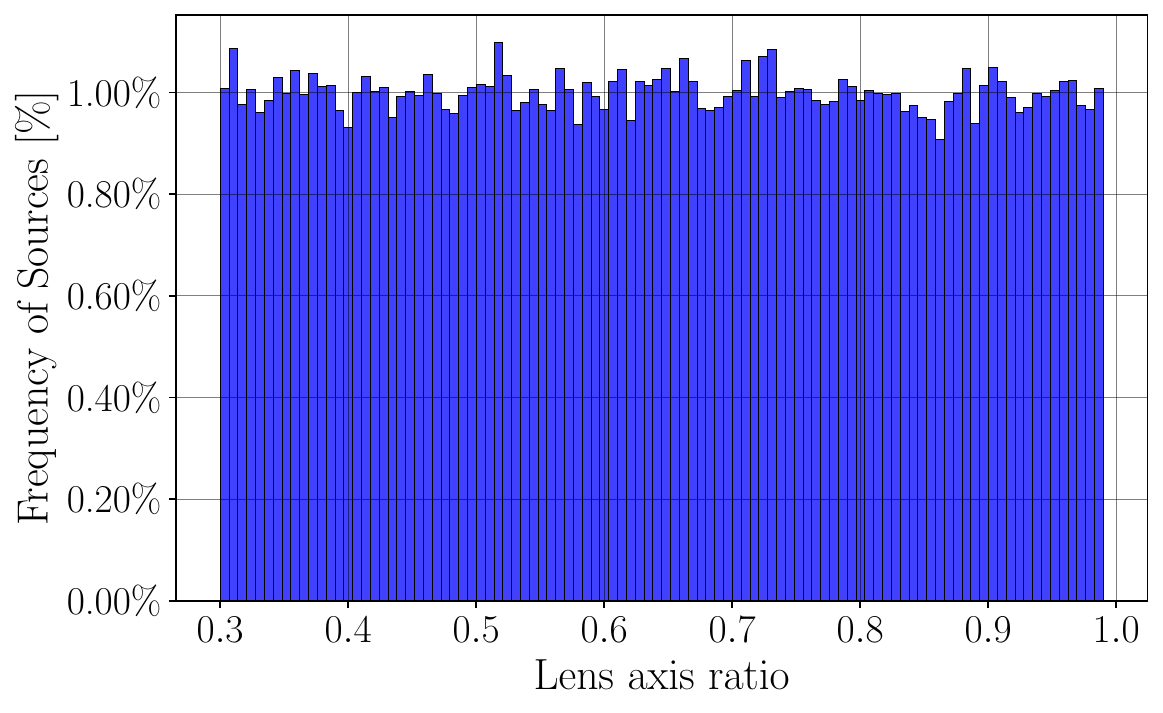}
       % \caption{Caption for Plot 1}
        \label{fig:sub1}
    \end{minipage}%
    \hfill
    \begin{minipage}{0.33\textwidth}
        \centering
        \includegraphics[width=\linewidth]{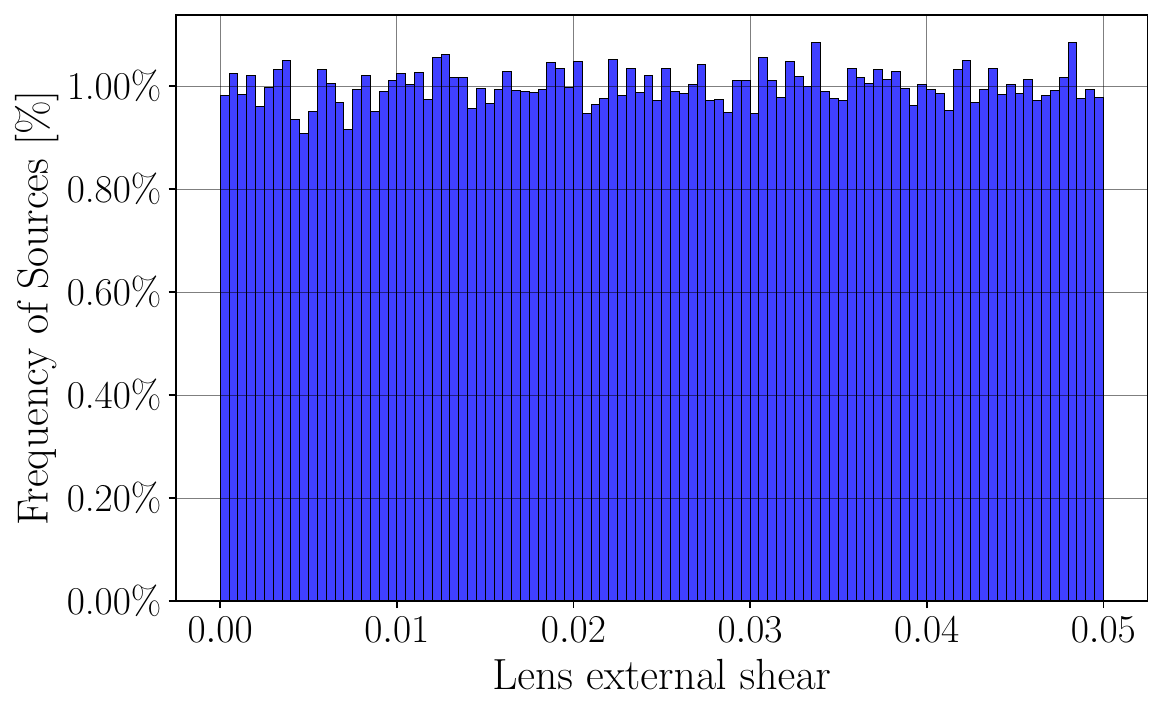}
        %\caption{Caption for Plot 2}
        \label{fig:sub2}
    \end{minipage}%
    \hfill
    \begin{minipage}{0.33\textwidth}
        \centering
        \includegraphics[width=\linewidth]{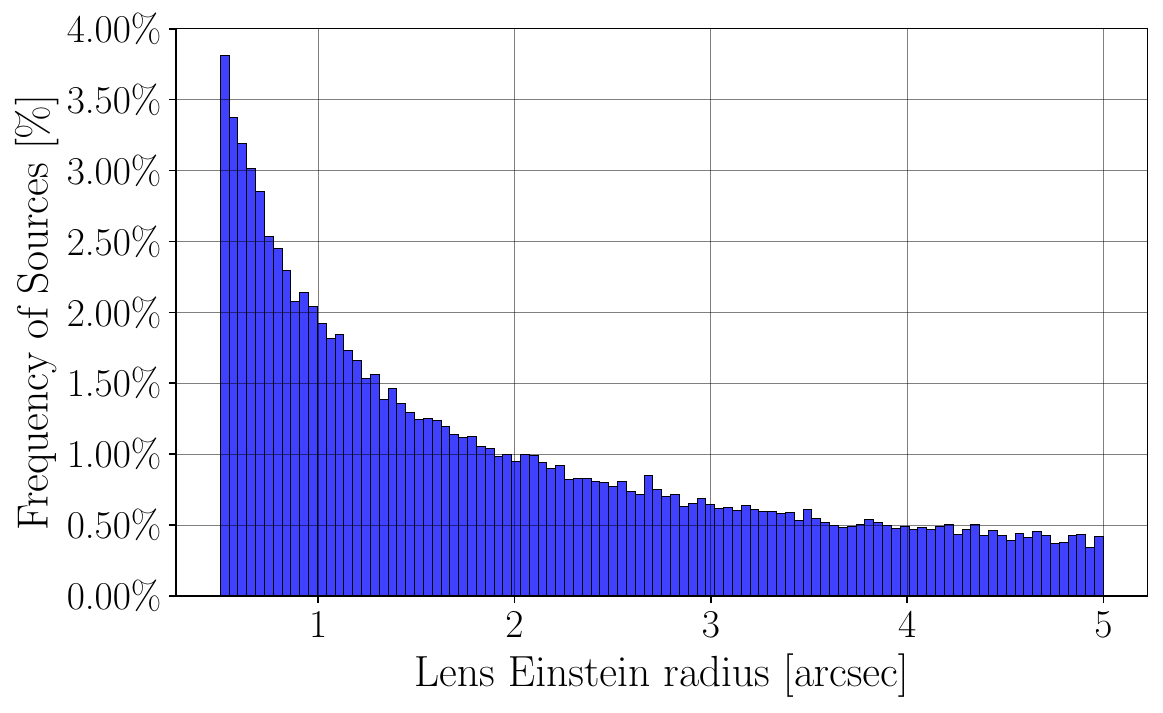}
        %\caption{Caption for Plot 3}
        \label{fig:sub3}
    \end{minipage}

    \caption{The distribution of the lens model parameters used for the training dataset; these are (left panel) the lens axis ratio (b/a), (middle panel) the lens external shear ($\gamma_{\rm ext}$ ), and (right panel) the lens Einstein radius ($\theta_{\rm E}$ ). Notably, the Einstein radii follow a logarithmic distribution, while the other parameters adhere to a flat distribution. The position angles of the ellipsoidal mass distribution and the external shear were set randomly between $\pm 90$ deg.}
    \label{fig:mock_lens_params}
\end{figure*}

We model the sources using a S\'ersic profile, and we exclude highly elliptical sources by restricting to axis ratios greater than 0.3 \citep{Petrillo2017}. The effective radius is randomly assigned within the range of 0.2 to 0.6 arcsec, while the S\'ersic index is randomly selected between 0.5 and 5, which extends lower when compared to the range considered by \citet{Petrillo2017}. This is done in order to consider a wider range of source morphologies in our samples. The range of S\'ersic indices and effective radii for the source galaxies shows a slight bias toward early-type galaxies \citep{he2020deep}. The major-axis position angle is also randomly assigned across the entire range of 0 to 180 deg, ensuring a robust and realistic dataset for source modelling. As evident from the mock lens parameter distribution shown in Fig.~\ref{fig:mock_lens_params}, the objective is not to replicate a statistically accurate representation of the real lens population. Instead, the emphasis is on densely populating the training dataset within the considered parameter space. This strategic approach, which has been previously employed by other studies \citep{Petrillo2019, Rezaei2022}, empowers developed model architectures to learn various configurations, even those that may be rare or currently unknown in real distributions. This approach enhances the model's ability to generalize beyond common scenarios.

In total, \(10^6\) mock lenses are generated, each covering a $20 \times 20$ arcsec field, introducing heightened complexity in both source and lens planes. Among these, 800,000 are randomly selected for the training phase, while the validation and test datasets each encompass 100,000 unique lens configurations.

\begin{table}
  \caption{The parameter ranges for the Singular Isothermal Ellipsoid (SIE) representing the lens and the S\'ersic model representing the source used in our simulations. These parameters contribute to the diversity of lensed images generated for training and testing purposes. The units and the range of values for each parameter are also indicated for reference.}
\begin{center}
\begin{tabular}{l l c}
Parameter              & Range & Unit \\
\hline
\multicolumn{3}{c}{Lens (SIE)}\\
\hline
Einstein radius      & 0.5 -- 5.0 & arcsec\\
Axis ratio           & 0.3 -- 1.0  & -\\
Major-axis angle     & 0.0 -- 180 & deg\\
External shear       & 0.0 -- 0.05 & -\\
External-shear angle & 0.0 -- 180 & deg\\
\hline
\multicolumn{3}{c}{Source (S\'ersic)}\\
\hline
Effective radius     & 0.2 -- 0.6 & arcsec\\
Axis ratio           & 0.3 -- 1.0 & -\\
Major-axis angle     & 0.0 -- 180 & deg\\
S\'ersic index       & 0.5 -- 5.0 & -\\
\hline
\end{tabular}

 \label{tab:sim_params}
	\end{center}
\end{table} 

\subsubsection{Creating real-looking lens systems}

To create realistic lens systems, we employ the method described by \citet{Petrillo2017}, which combines a chosen mock lens (detailed in Section~\ref{mock_lens}) with a potential lensing galaxy (LRGs; as explained in Section~\ref{lensing galaxy}). When combining a LRG and mock lensed emission, we adjust their peak brightness using a scaling factor \( ( 0.03 \leq K \leq 0.5) \), ensuring the lower magnitudes typically observed in lensing features relative to LRGs are preserved. Specifically, we scaled the brightness of the mock lens to \(K\) times the peak brightness of the selected LRG, allowing it to resemble the lensing galaxy in the system. Additional steps to create realistic lens systems include clipping negative pixel values to zero, which eliminates non-physical intensities, and applying a square-root stretch to emphasize low-brightness features, such as extended gravitational arcs. Finally, the images are normalized to a pixel value range of 0 and 1, ensuring uniformity across the dataset and optimizing the training process for the CNN.

\begin{figure}
    \centering
    \includegraphics[width=\columnwidth, trim=1cm 0cm 2cm 0cm, clip]{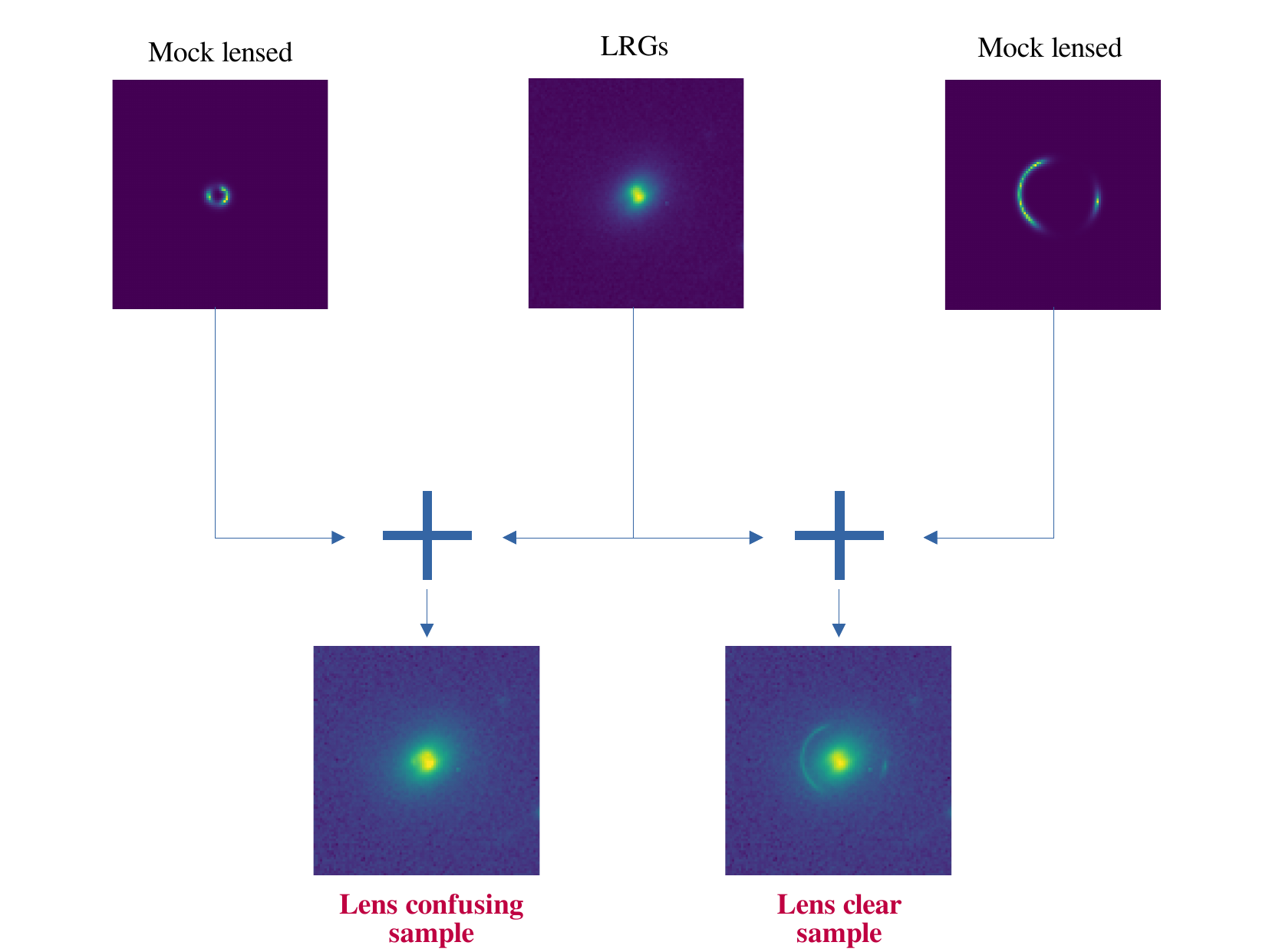}
    \caption{An example of a potential issue arising when the lensed emission is faint with respect to the brightness of the foreground lensing galaxy, and has an Einstein radius that is significantly lower than the effective radius of the foreground lensing galaxy. Two scenarios employing the same LRG as the foreground galaxy are shown, but with different lens configurations. As depicted, when lensed emission with a smaller Einstein radius is introduced to the selected LRG, the ring configuration of the lensed emission is entirely obscured by the LRG emission. The inclusion of lensed samples, such as the example on the left (labeled as 1 or lensed), may confuse the CNN model, as this sample resembles a non-lensed sample (labeled as 0) in our training dataset.}
    \label{fig:confusing_radii}
\end{figure}

Previous studies, such as by \citet{Petrillo2017}, have often employed a random selection strategy to pair mock lenses and foreground galaxies; however, in our investigation of the influence of training data on CNN performance, we identified potential limitations in this approach. Fig.~\ref{fig:confusing_radii} illustrates an example in which such a random strategy could cause potential problems by generating confusing training samples. Two scenarios are depicted: in one, a mock lens with a small Einstein radius is added to a LRG, resulting in a lens sample where the ring configuration of the lens is entirely hidden by the LRG emission. Conversely, incorporating a mock lens with a larger Einstein radius offers a more suitable match for the same LRG. This observation aligns with the equation \( M = \pi \rho \theta^2 \), where $\rho$ is the average surface mass density inside the Einstein radius (\( \theta \)). It is proportional to the square root of the enclosed mass (\( M \)). Assuming the enclosed mass correlates with the total integrated surface brightness, the Einstein radius is linked to the total flux of a galaxy. 

Avoiding the generation of samples similar to the one on the left side of Fig.~\ref{fig:confusing_radii} enhances quality assurance for the CNN. This sample, which would be labeled as lensed despite presenting no clear lensing emission, can be considered a non-lensed sample. In other words, incorporating nearly identical samples with differing labels into the training dataset leads to confusion for the model. This underscores the importance of carefully selecting the pair of foreground galaxy and mock lensed emission, as it significantly influences the composition of the generated lens population in the training dataset.

By avoiding such confusing samples, the model trains more effectively and produces the expected output, translating to a lower false positive (FP) rate. Reducing the FP rate is particularly important in large-scale surveys such as KiDS and Euclid, where the volume of data is immense, and manual inspection of each candidate lens is impractical. High FP rates make it challenging to accurately identify true lenses. This not only wastes valuable time and resources, but also introduces uncertainties that can propagate into subsequent analyses, leading to erroneous conclusions about the properties and distribution of galaxies and dark matter. 

\section{Method}\label{method}
In this section, we provide an overview of the architecture employed in our deep learning algorithm designed for the detection and ranking of gravitational lensing candidates. CNNs recognized for their adeptness in processing input imaging data with a topological structure, stand out as the primary approach for object detection and classification. The strength of CNNs lies in their capacity to utilize multiple layers, each serving a distinct function, and the arrangement of these layers can produce various convolutional components. As a consequence, the efficacy of a CNN is closely tied to the specific components implemented, and the performance may vary based on the application's requirements. 

To quantify the performance of our network in terms of dissimilarity between estimated and true class labels, we employ loss functions. In our task of binary classification to distinguish between lensed and non-lensed objects, we have chosen the Binary Cross Entropy (BCE) form of the loss function for our evaluations. The BCE loss function, represented as,
\begin{equation}
{\rm BCE} = -\frac{1}{N} \sum_{i=1}^{N} \left[ y_i \log p_i + (1 - y_i) \log (1 - p_i) \right],
\end{equation}
is employed where \( y_i \) denotes the given class label for the \( i \)th sample in our dataset of \( N \) training samples, and \( p_i \) represents the estimated probability of the model indicating the \( i \)th sample as a strong gravitational lens system.

While some studies have developed custom architectures for strong lens detection \citep{Rezaei2022}, leveraging existing network structures is a common practice in the field. Among the widely utilized architectures, \textit{ResNet} \citep{He2016} stands out as one of the most popular choices in the strong lensing literature, as evidenced by studies such as from \citet{Petrillo2017} and \citet{Lanusse2018}.  The \textit{ResNet} model is built on the concept of training deeper CNNs by incorporating shortcuts or by skipping connections between the front and back layers. This strategy helps in facilitating the backpropagation of gradients during training, allowing for better optimization of the model. In a comparative analysis conducted by \citet{Nagam2023}, the performance of \textit{DenseNet} \citep{Huang2017} in comparison to \textit{ResNet} was assessed. The findings revealed that \textit{DenseNet} achieved comparable true positive rates while exhibiting lower false positive rates. The \textit{DenseNet} model builds upon the skipped connections concept, but introduces dense connections between all previous and subsequent layers. 

This unique characteristic allows \textit{DenseNet} to achieve superior performance compared to \textit{ResNet}, all while requiring fewer parameters and incurring less computational cost. Motivated by those results, we have opted for \textit{DenseNet} in our further analysis. Specifically, we explore multiple variants, including \textit{DenseNet-121} and \textit{DenseNet-169}.

Furthermore, we investigate various architectures based on the understanding that the selection of model architecture plays a crucial role in determining the overall performance in tasks such as strong lens detection.
Taking this direction further, we consider another branch of CNN architectures, called  \textit{EfficientNet} \citep{tan2019efficientnet}, which achieve state-of-the-art performance on image classification tasks while also being computationally efficient. In particular, we investigate various versions, such as \textit{EfficientNet-B3} and \textit{EfficientNet-B4}. This comprehensive analysis aims to uncover insights into how different neural network architectures influence the accuracy and reliability of strong lens detection models. In the following, we provide an overview of these selected architectures. 

\subsection{\textit{DenseNet}}
\textit{DenseNet}, short for Densely Connected Convolutional Networks, is a type of neural network architecture that emphasizes dense connectivity within "dense blocks". In each dense block, every layer receives the feature maps generated by all preceding layers as input, while also passing on its own feature maps to every subsequent layer in the block. This dense connectivity structure is distinct from traditional CNNs, where each layer is only connected to the immediately following layer. By establishing direct connections between all layers in a block, \textit{DenseNet} allows each layer to directly access the features of all previous layers, promoting both efficient information flow and rich feature representation. \textit{DenseNet} comes in various versions, such as \textit{DenseNet-121}, \textit{DenseNet-169}, and \textit{DenseNet-201}, where the numbers represent the total number of layers in each network. The choice of model variant depends on the complexity of the task and the available computational resources. For tasks that demand a balance between model complexity and computational efficiency, \textit{DenseNet-121} and \textit{DenseNet-169} are widely adopted due to their favorable trade-off between performance and resource consumption. We refer the interested reader to the review by \citet{Huang2017} for a detailed discussion on the architecture of \textit{DenseNet}.

\subsection{\textit{EfficientNet}}

\textit{EfficientNet} is a family of CNNs with the key innovation of compound scaling, which enables an optimal trade-off between model size and performance by uniformly scaling the network's depth, width, and resolution. \textit{EfficientNet} has been widely used in a variety of scientific applications, such as galaxy morphology classification, \citep{kalvankar2020galaxy}, spectral classification of astronomical objects \citep{wu2023automatic}, skin cancer detection \citep{VENUGOPAL2023100278}, brain tumor detection \citep{nayak2022brain}, and lung cancer detection
\citep{raza2023lung}. 

The key innovation of \textit{EfficientNet} lies in the balance between model depth, width, and resolution, as governed by a compound scaling method. This approach ensures that the network scales efficiently across these dimensions, making it well-suited for diverse tasks. Each variant of \textit{EfficientNet} is denoted by a scaling factor (e.g., \textit{EfficientNet-B3}, \textit{EfficientNet-B4}), reflecting its capacity for increased depth and width. These scaling factors allow users to choose a model that aligns with the specific requirements of their task and computational resources. This concept can be mathematically expressed as,
\begin{equation}
    \alpha .\, \beta^2.\, \gamma^2 \approx 2
\end{equation}
for $\alpha \geqslant 1$, $\beta \geqslant 1$, and $\gamma \geqslant 1$. Here, the depth is $\alpha^{\phi}$, the width is $\beta^{\phi}$ and the resolution is $\gamma^{\phi}$, where $\phi$ denotes the scaling coefficient that uniformly scales the network.
The choice of scaling coefficients impacts the trade-off between model complexity and computational efficiency, making \textit{EfficientNet} a versatile architecture that is adaptable to different resource constraints and task requirements. Further details on the structure of \textit{EfficientNet} are given by \citet{tan2019efficientnet}. 

 In our investigation, we incorporate \textit{EfficientNet-B3} and \textit{EfficientNet-B4}, considering their widespread adoption and ability to strike a suitable balance between model complexity and computational efficiency in various applications. The initial weights were randomly set using a uniform distribution. While different weight initializations can impact the training process during the early epochs, our experiments show that the network converges shortly after this period. Thus, the observed performance remains largely unaffected by the initial weight settings.

\subsection{Evaluation criteria}

To assess and compare the effectiveness of different methodologies and training data strategies on the test dataset, it's essential to establish appropriate evaluation criteria. Given that lens detection is treated as a classification problem, where samples are categorized as lensed or non-lensed, the following representation can be utilized for evaluation purposes:
\[
\begin{array}{cc|cc}
\multicolumn{2}{c}{} & \multicolumn{2}{c}{\text{True Data}} \\
\cline{3-4}
\multicolumn{2}{c|}{} & \text{Lens} & \text{Not Lens} \\
\cline{2-4}
\text{Test Results} & \text{Lens} & TP & FP \\
\cline{2-4}
 & \text{Not Lens} & FN & TN \\
\cline{2-4}
\end{array}
\label{tab:sample_confusion}
\]

In this representation, True Positive (TP) indicates correctly identified gravitational lens systems, True Negative (TN) corresponds to accurately recognized non-lensed sources, False Positive (FP) denotes mis-classification of non-lensed sources as gravitational lenses, and False Negative (FN) refers to gravitational lensing events missed by the algorithm and classified as non-lensed sources.

Based on these terms, several evaluation criteria can be defined. Accuracy measures the proportion of correctly identified samples (TP and TN) out of the total number of samples. Precision quantifies the ratio of true positives to the sum of true positives and false positives, indicating the reliability of positive predictions. Recall assesses the fraction of true positives correctly identified by the algorithm, reflecting the model's completeness. Fall-out, also known as the false positive rate, calculates the proportion of negative samples incorrectly classified as positive, providing insight into the purity of detected lens candidates. Mathematically, these metrics are represented as:
\begin{equation}
\begin{aligned}
\text{Accuracy} &= \frac{TP + TN}{TP + FN + TN + FP}, \\
\text{Precision} &= \frac{TP}{TP + FP}, \\
\text{Recall} &= \frac{TP}{TP + FN}, \\
\text{Fall-out or FP rate} &= \frac{FP}{FP + TN}.
\end{aligned}
\end{equation}
In assessing gravitational lens search algorithms, the emphasis extends beyond completeness alone, especially given the anticipation of a large number of gravitational lenses to be detected with upcoming all-sky surveys. Rather, the focus often centres on achieving a low FP rate, with the aim of identifying a high number of genuine lens candidates in the ranked list. Additionally, the Receiver Operating Characteristic (ROC) curve serves as another valuable metric. The ROC plot visually represents the trade-off between the TP rate and the FP rate for each model. Each point on the ROC curve signifies a different threshold for classifying samples as positive or negative based on their predicted probabilities. By examining the ROC plot, we can discern how well each model discriminates between positive (lensed) and negative (non-lensed) samples. A model with better performance will exhibit a curve that closely approaches the top-left corner of the plot, indicating higher TP rate and lower FP rate across various threshold values. The comparison of ROC curves for different models provides insights into their relative effectiveness in identifying lensed samples. This analysis aids in selecting the most suitable model for the task at hand, considering both sensitivity to TPs and robustness against FPs.

\section{Results} \label{results}
The success of any machine learning model relies on the quality of the training dataset. This section examines various properties of training datasets, focusing on their impact in strong lens detection projects. The relevance and representation of the training dataset are crucial. A well-curated dataset must encompass a diverse and representative set of examples that reflect the variety and complexity of real-world data. In the context of strong lens detection, this means including images with different types of strong lensing phenomena, as well as a high variety in the non-lensed samples. Also, a representative dataset helps the model learn to generalize from the training data to unseen data, improving its robustness and accuracy. Another important factor is accurate labeling as the model relies on these labels to learn the correct associations between input data and the desired output. An example of the importance of accurate labeling in the context of strong lens detection is provided in Fig.~\ref{fig:confusing_radii}.

Considering the significant impact of training data on the performance of CNNs, we have provided two novel strategies to handle the training dataset, complementing the conventional approach typically adopted in the literature (e.g. \citealt{Petrillo2019,Nagam2023}), which we refer to as the Vanilla strategy. Our aim is to craft a dataset that enhances the purity of detected candidates. While completeness is still a consideration, in this study, we prioritize the necessity of mitigating FPs.

Our first approach, termed "Applied1", prioritizes the treatment of non-lensed samples within the training dataset. This strategy is tailored to address specific challenges associated with the characterization of non-lensed objects. Conversely, our second approach, "Applied2", focuses on optimizing the representation of the lens population within the training dataset, thereby enhancing the CNN's ability to accurately identify and classify gravitational lensing events. Further elaboration on these strategies is provided below, including their respective methodologies and rationales for effectively training CNNs in gravitational lens detection tasks.

Although we have tested CNN architectures on various training datasets, we ensured that each round incorporated the same number of samples to provide a fair analysis of model performance. Each CNN architecture was trained on a total of 500,000 samples. This training dataset was balanced with 250,000 samples labeled as lensed and 250,000 as non-lensed.

\subsection{Vanilla} \label{van}

\begin{figure*}
    \centering
    \includegraphics[ scale=0.56]{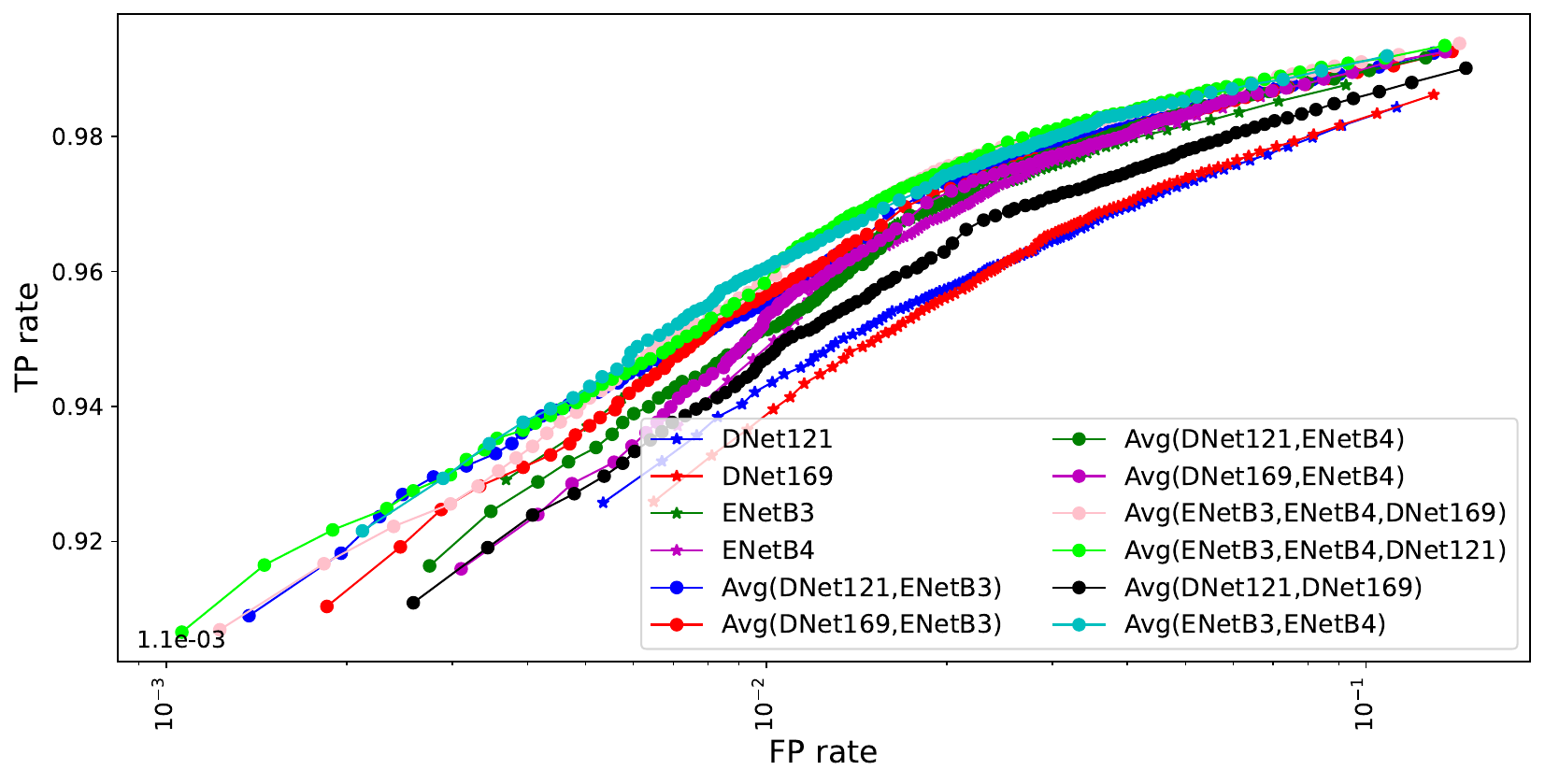}
    \caption{The ROC plot demonstrates how different machine learning models, trained with the Vanilla setting, perform in distinguishing between positive (lensed) and negative (non-lensed) samples across various thresholds balancing true positive (TP) and false positive (FP) rates. For presentation purposes, the range of the ROC plot has changed from [0,1] to the current display. By comparing these curves, we can identify the most efficient model for detecting lensed samples, while minimizing FPs. The plotted results show the improvement in FP rate as we use an ensemble technique, by averaging the predicted lens probability of individual models. The best performing ensemble belongs to averaging the output of \textit{EfficientNet-B3}, \textit{EfficientNet-B4} and \textit{DenseNet-121} with the FP rate of $1.1 \times 10^{-3}$ and a TP rate of 0.906.}
    \label{fig:roc_vanila}
\end{figure*}

\begin{table}
    \caption{The main evaluation metrics, such as true positive (TP) and false positive (FP) rates, derived from training the featured models, using the Vanilla setting on the training dataset. The predicted lensing probability from each model is a value between 0 and 1. However, for calculating the evaluation metrics, we have established a threshold of 0.99 to differentiate between lensed and non-lensed samples. The test dataset consists of 96,072 samples equally distributed between lensed and non-lensed categories.}
    \centering
    \begin{tabular}{ccc}
    Model & TP & FP\\
    \hline
       DenseNet-121  &0.925& $5.3 \times 10^{-3}$ \\
%       \hline
       DenseNet-169 &0.925&$6.4 \times 10^{-3}$ \\
%       \hline
       EfficientNet-B3 &0.929&$3.6 \times 10^{-3}$ \\
%       \hline
       EfficientNet-B4 &0.937&$7.1 \times 10^{-3}$ \\
%       \hline
       DenseNet-121, EfficientNet-B3 &0.908&$1.3 \times 10^{-3}$ \\
%       \hline
       DenseNet-169, EfficientNet-B3 &0.910 &$1.8 \times 10^{-3}$ \\
%       \hline
       DenseNet-121, EfficientNet-B4 &0.916&$2.7 \times 10^{-3}$ \\
%       \hline
       DenseNet-169, EfficientNet-B4 &0.915&$3.1 \times 10^{-3}$ \\
%       \hline
       EfficientNet-B3, EfficientNet-B4 &0.921&$2.1 \times 10^{-3}$ \\
%       \hline
       DenseNet-121, DenseNet-169 &0.910 &$2.7 \times 10^{-3}$ \\
%\hline
    DenseNet-121, EfficientNet-B3, EfficientNet-B4&0.906 &$1.1 \times 10^{-3}$\\
%    \hline
    DenseNet-169, EfficientNet-B3, EfficientNet-B4&0.906 &$1.2 \times 10^{-3}$\\
    \hline
    \end{tabular}

    \label{tab:roc_vanilla}
\end{table}

\begin{figure}
\begin{center}
\begin{minipage}[b]{\columnwidth}
\centering
\includegraphics[height=2.85cm]{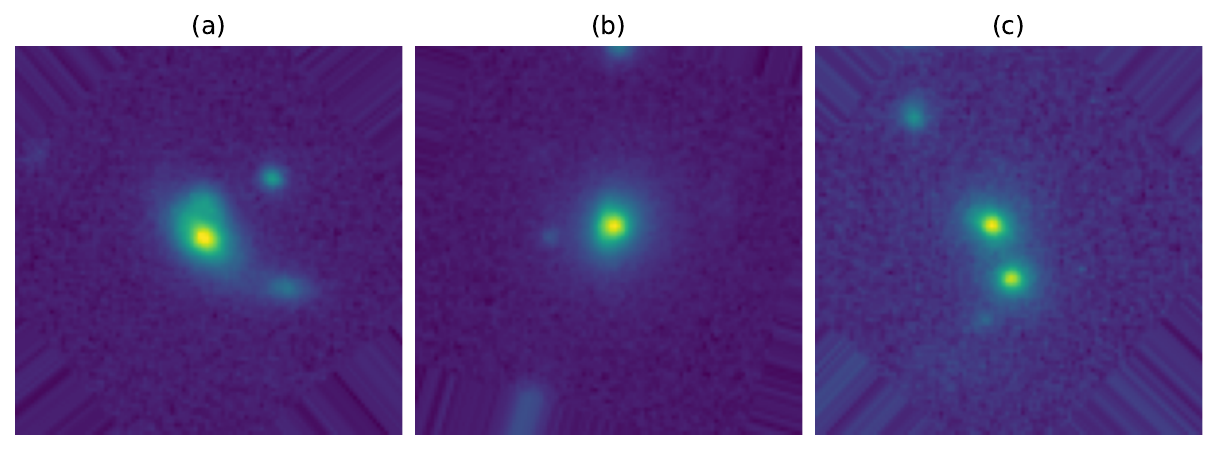}
\end{minipage}
\begin{minipage}[b]{\columnwidth}
\centering
\includegraphics[height=2.85cm]{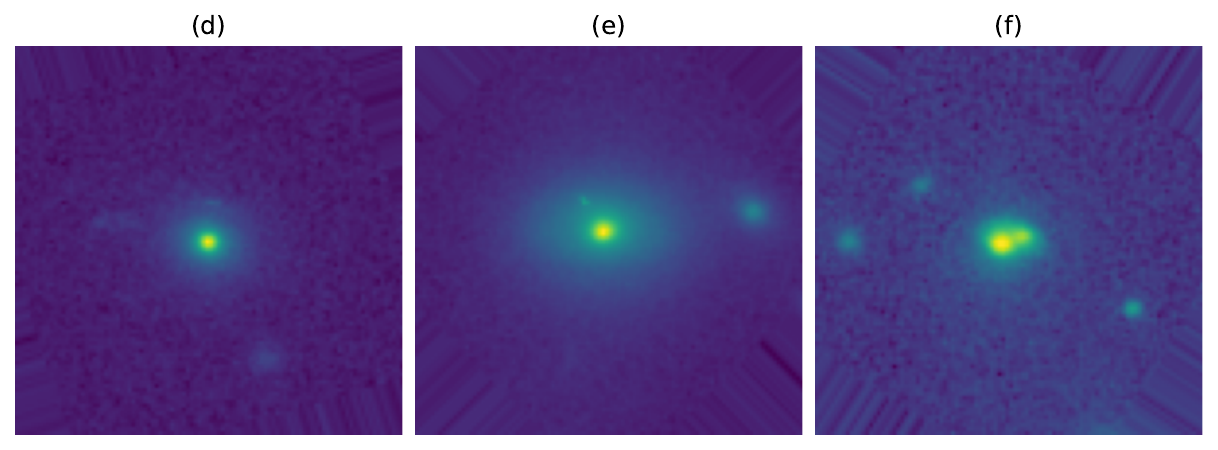}
\end{minipage}
\newpage
\caption{A selection of LRGs falsely detected as strong lens systems by the Vanilla setting. These samples (FPs) are having a high complexity value of 200 or more, which are not well represented in the Vanilla setting. More details regarding these samples are provided in Table~\ref{tab:high_comp_details}.}
\label{fig:sample_high_compactness}
\end{center}
\end{figure}

The Vanilla setting, pioneered by \citet{Petrillo2017} and recently utilized by \citet{Nagam2023} with an expanded repertoire of non-lensed samples, stands as a benchmark methodology for gravitational lens detection tasks using KiDS data. As described above, the generation of lensed samples entails a random pairing of a lens and an LRG sample to create a realistic-looking strong lensing sample. Within this framework, non-lensed samples are predominantly drawn from a pool of observed spiral (contaminations) and elliptical galaxies. These samples, originating from authentic KiDS observations, undergo augmentation to mitigate overfitting and enhance model performance. While \citet{Nagam2023} maintains a ratio of 20 per cent LRGs to 80 per cent contamination samples in their non-lensed selection process, we have opted for a balanced approach, incorporating an equal number of LRGs and contaminations in our dataset. This deliberate choice aims to provide the CNNs with a more comprehensive representation of LRGs, thereby strengthening their capacity to differentiate between LRGs acting as foreground galaxies in gravitational lens systems and those exhibiting typical LRG characteristics devoid of lensing effects.

Fig.~\ref{fig:roc_vanila} shows the performance comparison of the four different models: \textit{DenseNet-121}, \textit{DenseNet-169}, \textit{EfficientNet-B3}, and \textit{EfficientNet-B4}. Since the model output represents the probability of an object being a lens, ranging between 0 and 1, setting a threshold is necessary to classify objects into two classes: lensed and non-lensed. To ensure a stringent selection of strong lenses, a threshold of 0.99 is set for further analysis. This decision is driven by our primary goal of reducing the FP rate. As illustrated in Fig.~\ref{fig:roc_vanila}, increasing the threshold does not significantly impact the TP rate. In other words, it is acceptable to sacrifice a small percentage of the TP rate to achieve a better FP rate.

Building upon findings of \citet{Rezaei2022, Andika2023}, it has been observed that averaging the output probabilities of multiple models can effectively mitigate FP rates. This improvement stems from the models' tendency to disagree on suspicious cases, thus allowing for a voting scheme to be applied. As depicted in Fig.~\ref{fig:roc_vanila}, averaging the model outputs indeed yields enhanced results. 

\cref{tab:roc_vanilla} provides a detailed overview of each method's performance when the threshold is set at 0.99. Notably, the average of \textit{EfficientNet-B3}, \textit{EfficientNet-B4}, and \textit{DenseNet-121} demonstrates the most promising outcomes, with a FP rate of $1.1 \times 10^{-3}$ and an acceptable TP rate of 0.90. Following closely, the average of \textit{EfficientNet-B3}, \textit{EfficientNet-B4}, and \textit{DenseNet-169} achieves a similar TP rate and a slightly higher FP rate of $1.2 \times 10^{-3}$. These results underscore the effectiveness of leveraging ensemble methods to enhance the performance of gravitational lens detection algorithms. 

In order to comprehensively understand the model behavior and analyze its performance, we conducted a thorough investigation into the samples contributing to the FP rate of $1.1 \times 10^{-3}$ by our best model. Several examples are provided in Fig.~\ref{fig:sample_high_compactness} and their properties are given in Table~\ref{tab:high_comp_details}. Our analysis revealed that a significant portion of these FPs stem from the misclassification of extended LRGs as strong gravitational lensed samples. This observation, coupled with the complexity distribution shown in the lower panel of Fig.~\ref{fig:lensingGalaxyInfo}, indicates that our training data for LRG samples lacks a sufficient number of instances representing LRGs with extended emission or those with nearby contaminants. This realization prompted us to introduce the first variation of this study, which is detailed in the next sub-section.

This investigation highlights the importance of ensuring the diversity and representativeness of training data not only in lensed samples, but also in non-lensed samples. By addressing the imbalance in the representation of extended LRGs in our training dataset, we aim to enhance the robustness of our model in distinguishing between genuine gravitational lenses and other astronomical objects.

\begin{table}
    \centering
     \caption{Properties of FPs detected by the best performing model using the Vanilla setting. The indices correspond to the images depicted in Fig.~\ref{fig:sample_high_compactness}. The lens probability and uncertainty metrics were derived from the averaged predictions of the lensing probability across the \textit{EfficientNet-B3}, \textit{EfficientNet-B4}, and \textit{DenseNet-121} models. These findings indicate that all three models consistently predict a probability exceeding 99 per cent for this selection of LRGs, even though they are non-lensed samples.}

    \begin{tabular}{cccc}
       Index &  Lens probability & Complexity & Effective radii\\
       \hline 

       a &  0.997  & 271 & 21.28\\ 

        b&  0.999 & 282 & 24.73\\ 

        c&  0.995 &283 & 32.41\\ 
        d&  0.999&  297 & 21.17\\ 

        e&  0.999 &  285 & 25.38\\ 

        f&  0.997 & 400 & 50.23\\ \hline

    \end{tabular}
       \label{tab:high_comp_details}
\end{table}

\begin{figure}
    \centering
    \includegraphics[width=\columnwidth]{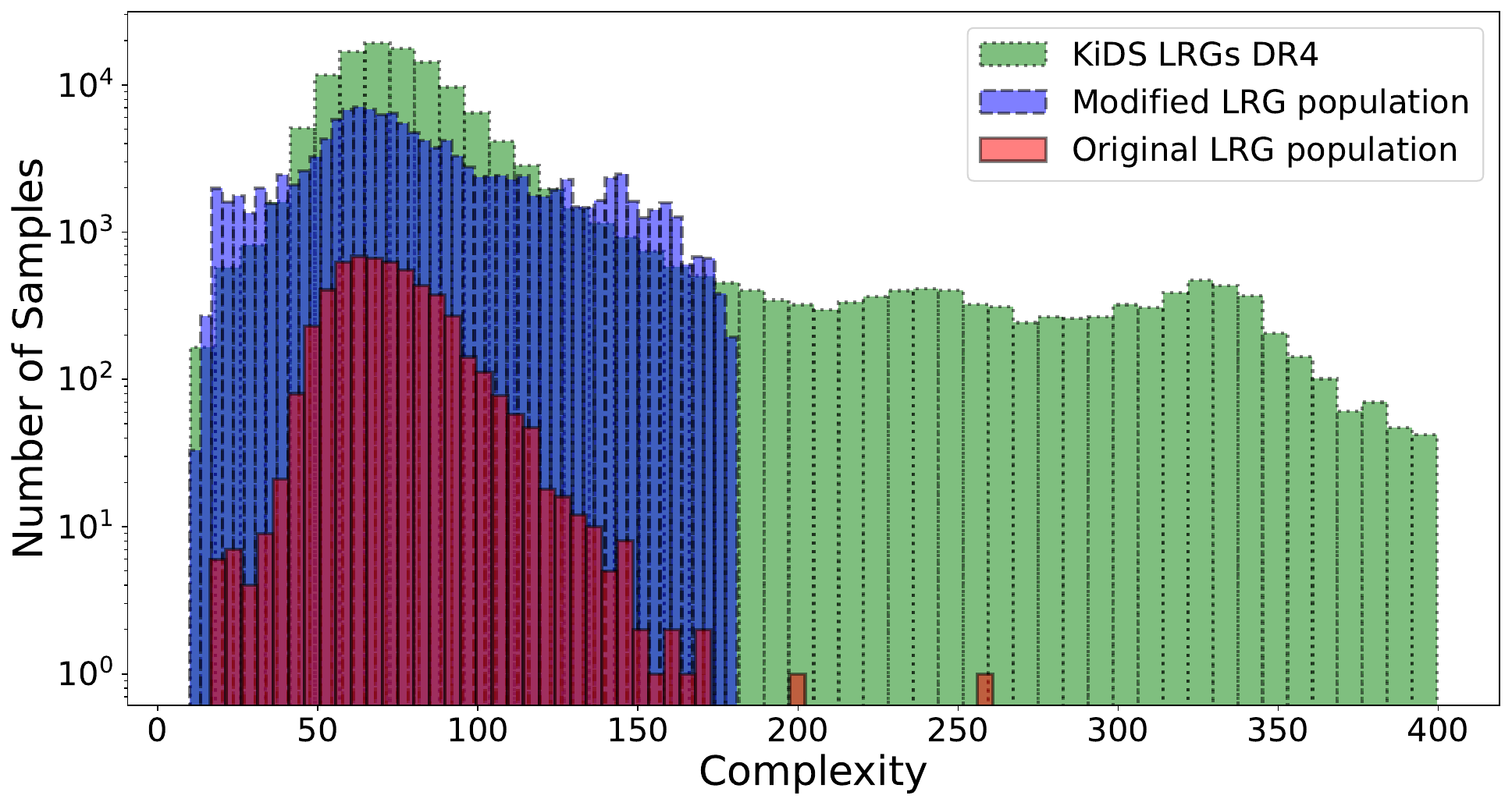}
    \caption{A comparison of the complexity between observational KiDS LRGs data provided by \citet{RuiLi2020} and the LRG samples within our non-lensed population from the training dataset. It highlights the adjustments made to the LRG population, achieved through strategic augmentation of LRG samples (in blue). These modifications aim to enhance the visibility of sources with less representative samples, particularly those with complexity values exceeding 100. By augmenting these samples, the CNN is exposed more to the complex LRGs and, thereby enhancing its ability in identifying and analyzing LRGs as non-lensed samples.}
    \label{fig:compactness}
\end{figure}

\subsection{Applied1} \label{applied1}
\begin{figure*}
    \centering
    \includegraphics[scale=0.6]{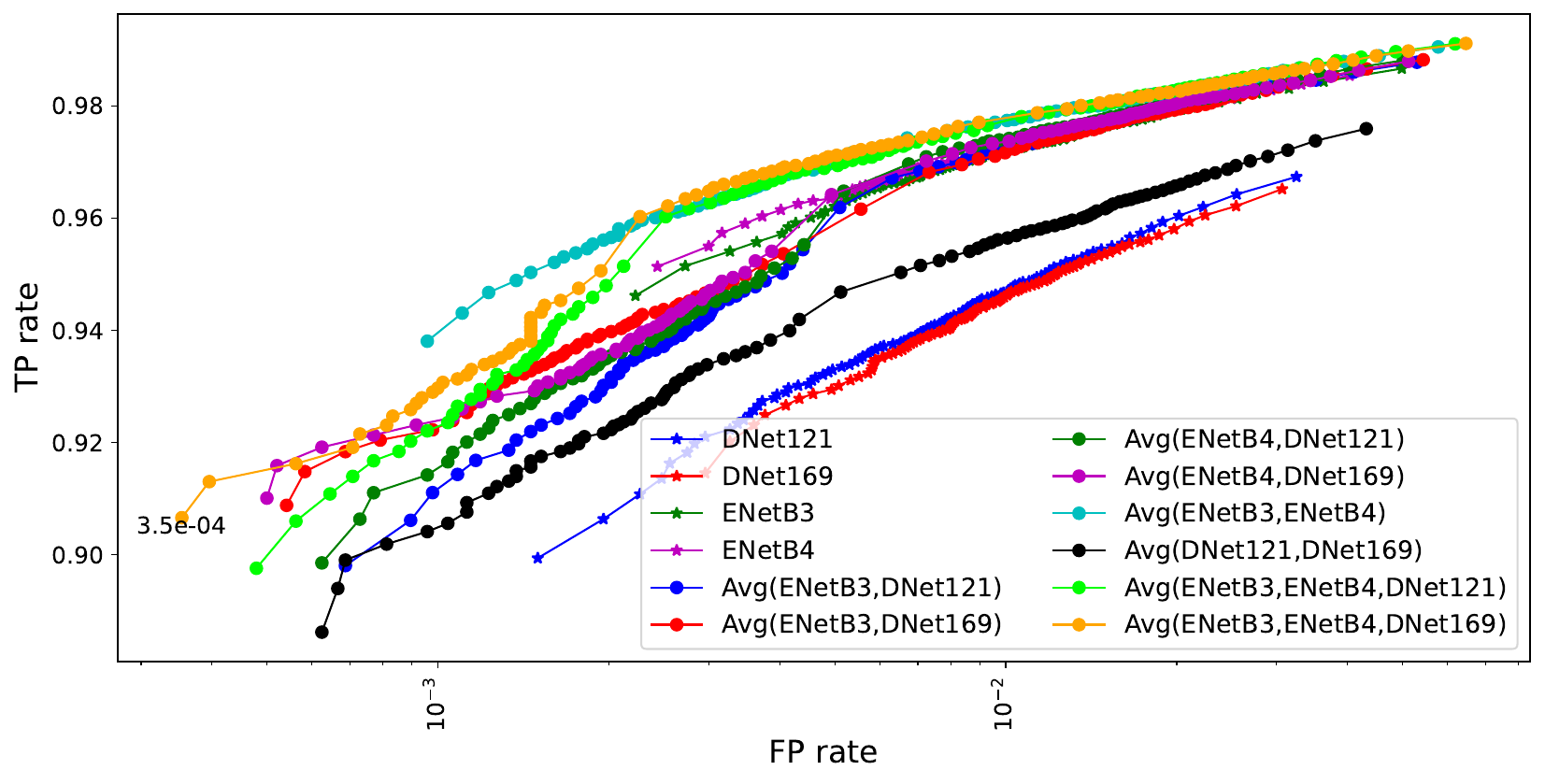}
    \caption{The ROC plot comparing the performance of trained CNN models under the Applied1 scenario in distinguishing between lensed and non-lensed samples. Given the significance of minimizing FPs, the most effective model is one that can provides a purer selection of candidates. Consequently, the ensemble comprising \textit{EfficientNet-B3}, \textit{EfficientNet-B4}, and \textit{DenseNet-169}, with an FP rate of $3.5 \times 10^{-3}$ and a TP rate of 0.906, emerges as the top-performing model.}
    \label{fig:roc_app1}
\end{figure*}

\begin{table}
    \centering
\caption{The primary evaluation metrics obtained from training the CNN models using the Applied1 setting on the training dataset. The models predict lensing probability values ranging between 0 and 1. However, to calculate the evaluation metrics, we have set a threshold of 0.99 to distinguish between lensed and non-lensed samples. The test dataset comprises 96,072 samples evenly split between lensed and non-lensed categories.}
    \begin{tabular}{ccc}
    Model & TP & FP\\
    \hline
       DenseNet121  &0.899&$1.4 \times 10^{-3}$ \\

       DenseNet169 &0.914&$2.9 \times 10^{-3}$ \\

       EfficientNetB3 &0.946&$2.2 \times 10^{-3}$ \\

       EfficientNetB4 &0.951&$2.4 \times 10^{-3}$ \\

       DenseNet121, EfficientNetB3 &0.898& $6.8 \times 10^{-4}$ \\

       DenseNet169, EfficientNetB3 &0.900&$5.4 \times 10^{-4}$ \\

       DenseNet121, EfficientNetB4 &0.898&$6.2 \times 10^{-4}$ \\

       DenseNet169, EfficientNetB4 &0.910&$4.9 \times 10^{-4}$ \\

       EfficientNetB3, EfficientNetB4 &0.938&$9.5 \times 10^{-4}$ \\

       DenseNet121, DenseNet169 &0.886&$6.2 \times 10^{-4}$ \\

    DenseNet121, EfficientNetB3, EfficientNetB4&0.897 &$4.7 \times 10^{-4}$\\

    DenseNet169, EfficientNetB3, EfficientNetB4&0.906 &$3.5 \times 10^{-4}$\\
    \hline
    \end{tabular}
    
    \label{tab:roc_app1}
\end{table}

\begin{table}
    \centering
     \caption{The lens probability of the FPs detected using the Vanilla setting that were not detected as FPs using the Applied1 scenario. Alongside the ensemble predicted lens probability, the individual lens probabilities for \textit{EfficientNet-B3} (ENet-B3), \textit{EfficientNet-B4} (ENet-B4), and \textit{DenseNet-169} (DNet-169) are provided. All of these predictions are based on the Applied1 settting. A visual representation of these samples can be found in Fig.~\ref{fig:sample_high_compactness}, while additional details regarding the Vanilla setting predictions are presented in Table~\ref{tab:high_comp_details}.}
    \begin{tabular}{ccccc}
       index & Ensemble &ENet-B3 &ENet-B4&DNet-169\\
       \hline
       a &  0.65 &  0.950 & 0.997 & $3.3 \times 10^{-6}$\\

        b&  0.96 & 0.999 & 1 & 0.89\\

        c&  0.64 & 0.950 & 0.98 & $1.5 \times 10^{-3}$, \\

        d&  0.98& 0.999 & 1  &0.95\\

        e&  0.67 & 0.999 & 0.999 & $1.1 \times 10^{-2}$, \\

        f&  0.70 & 0.991 & 0.998 &0.12\\ \hline

    \end{tabular}
   
    \label{tab:high_comp_details_app1}
\end{table}

We now demonstrate how the properties of the training dataset can significantly influence the achieved results and the overall performance of the model. This is done by changing the characteristics and sample population within the training dataset, while maintaining consistency in the test dataset. 
Drawing insight from our analysis of the FP samples presented in Section~\ref{van}, 
here we focus on improving the representation of the training dataset for non-lensed samples. By incorporating this solution, we have achieved compelling results, which we detail below. 

Fig.~\ref{fig:compactness} shows the distribution of complexity values among the LRG samples within our training dataset. Notably, this distribution exhibits non-uniformity, which is particularly evident in the complexity range of 120 and higher, where less than 1.4 per cent of the LRG samples are represented. To validate these findings, we compare them with the complexity distribution of a larger LRG dataset produced by \citet{RuiLi2020}. This dataset comprises low-redshift samples $(z \leq 0.4)$ selected based on their $r$-band magnitude, limited to values below 20. The complexity distribution of these LRG samples is also shown in Fig.~\ref{fig:compactness}. Remarkably, the comparison reveals that the distribution of LRG complexity in our training dataset does not accurately reflect the distribution observed in real observational data.

Even when augmentation techniques are applied to the current LRG samples, the resulting distribution tends to mirror the original distribution, which worsens the imbalance between compactness categories. Therefore, augmentation alone does not offer a viable solution to this issue. To address this challenge, we propose a strategic augmentation approach aimed at achieving a more balanced distribution of samples across various complexity bins. Specifically, we advocate for augmenting the less frequent sources with greater variations compared to the more densely populated regions in the complexity distribution. This strategic augmentation aims to ensure a semi-uniform distribution of samples across various complexity categories, thereby enhancing the representativeness of our training dataset. The distribution of the generated LRG population with this strategy is also shown in Fig.~\ref{fig:compactness}.

Following this, we adapted our training dataset to align with the adjusted LRG population. However, we have kept other training parameters, such as the total number of training data points, the distribution of lensed samples, the learning rate, and other hyperparameters consistent with the Vanilla setting. The results obtained from this training data configuration, referred to as "Applied1", are presented in Fig.~\ref{fig:roc_app1}. It is evident that the best achieved FP rate comes from the ensemble of \textit{EfficientNet-B3}, \textit{EfficientNet-B4}, and \textit{DenseNet-169}, with a value of $3.5 \times 10^{-4}$, marking a notable decrease from the FP rate of $1.1 \times 10^{-3}$ observed for the Vanilla setting. With a total of 48,036 non-lensed samples in our test dataset, the number of FP detections has decreased from 53 to 17, which represents more than a threefold reduction. This effect will likely be further magnified when dealing with a larger test dataset, such as the 126,884 LRGs from the KiDS DR4, as described by \citet{RuiLi2020}.

Moreover, this reduction in the number of FPs has not compromised the TP rate. The specific details of the FP and TP rates are presented in Table~\ref{tab:roc_app1}. In comparison to the Vanilla setting, where the ensemble of \textit{DenseNet-121}, \textit{EfficientNet-B3}, and \textit{EfficientNet-B4} yielded the best results, in this scenario, \textit{DenseNet-169} has replaced \textit{DenseNet-121} in the ensemble. The combination of \textit{EfficientNet-B3}, \textit{EfficientNet-B4} and \textit{DenseNet-169} achieves a TP rate that is 0.9 per cent better, while exhibiting a slightly higher FP rate of $1.2 \times 10^{-4}$,  when compared to the ensemble of \textit{EfficientNet-B3}, \textit{EfficientNet-B4} and \textit{DenseNet-121}.

Table~\ref{tab:high_comp_details_app1} offers some insight to the variability of individual model predictions and their influence on the ensemble probability. Also, the recorded samples are shown in Fig.~\ref{fig:sample_high_compactness} and their predicted lensing probabilities under the Vanilla setting are given in Table~\ref{tab:high_comp_details}. A comparison between Tables~\ref{tab:high_comp_details} and \ref{tab:high_comp_details_app1} highlights the better performance of the training strategy implemented in Applied1, when compared to the results using the Vanilla setting.

\subsection{Applied2} \label{Applied2}
\begin{figure*}
    \centering
    \includegraphics[scale=0.56]{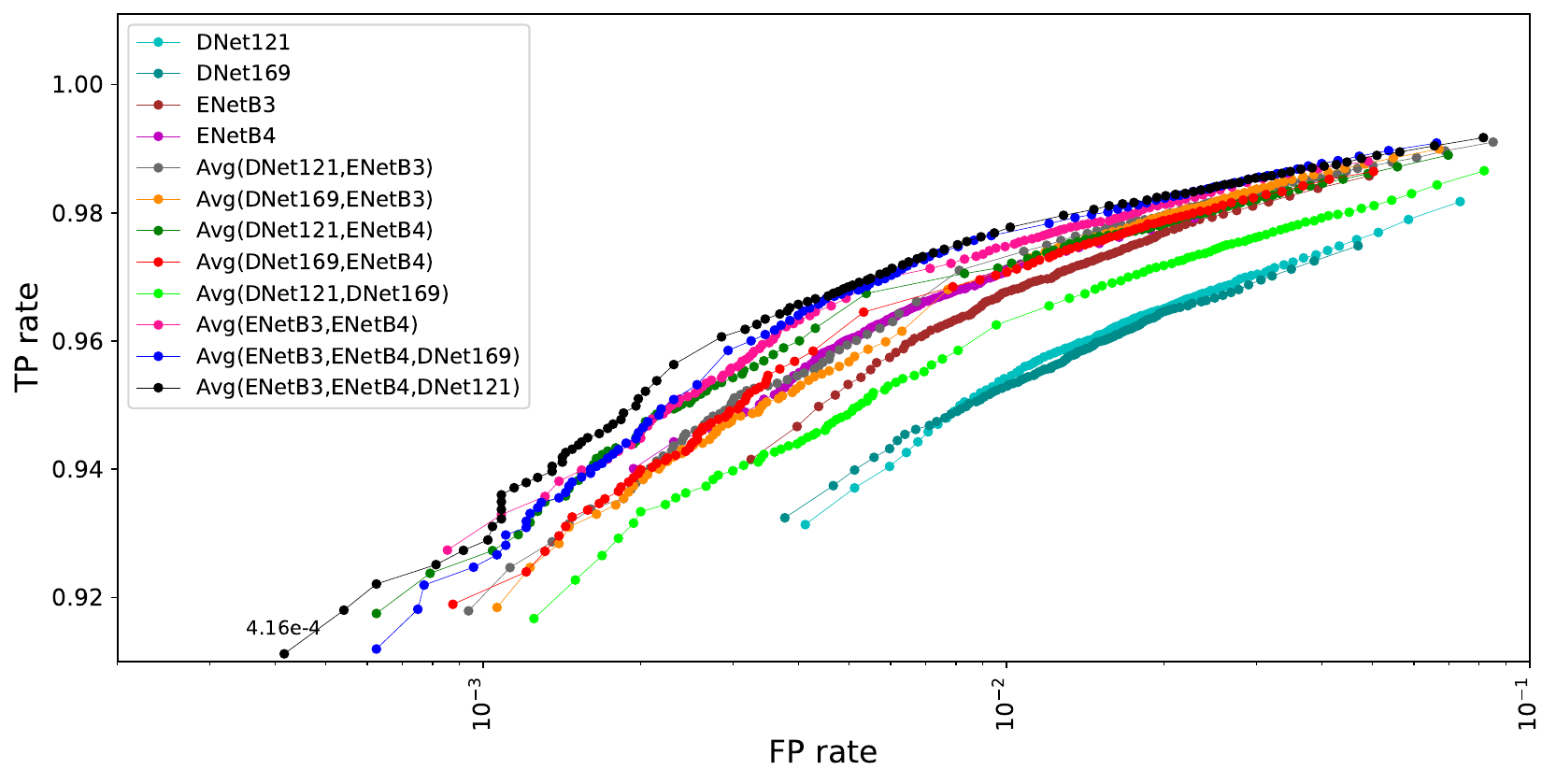}
    \caption{The ROC plot comparing the performance of trained CNN models under the Applied1 scenario in distinguishing between lensed and non-lensed samples. In this scenario, both the lensed and non-lensed populations have been altered compared to the Vanilla setting. The ensemble of \textit{EfficientNet-B3}, \textit{EfficientNet-B4}, and \textit{DenseNet-121} achieves a TP rate of 0.9112 with a FP rate of $4.16 \times 10^{-4}$. The achieved TP rate surpasses that of both the Vanilla and Applied1 settings.}
    \label{fig:roc_app2}
\end{figure*}

We now introduce our second modification to the training dataset. While the previous section addressed issues concerning the non-lensed population, the focus here is on potential challenges within the lensed population. As discussed previously in Fig.~\ref{fig:confusing_radii}, it is critical to consider the morphology of the foreground galaxy when pairing it with a mock lens. Our adjustments target how the lens population is constructed within the training dataset. As outlined in Section~\ref{training_data}, this process involves pairing mock lenses with foreground galaxies to simulate realistic strong gravitational lens systems. Through visual examination, we identified potential issues, particularly with small-separation lensed emission ($\theta_E < 0.85$ arcsec). An analysis of the mock lens parameters (see the right panel of Fig.~\ref{fig:mock_lens_params}) reveals that a considerable portion of existing samples in the training dataset belong to the small Einstein radii population, with approximately 23 per cent having Einstein radii below 0.85 arcsec.

Our primary focus is on selecting appropriate lensing galaxies, specifically tailored for these smaller Einstein radii lenses. An example illustrating the potential issue is presented in Fig.~\ref{fig:confusing_radii}, where an unsuitable choice of lensing galaxy results in a perplexing sample that lacks distinct lensed emission. Given the considerable likelihood (approximately 23 per cent) of such mock lenses being incorporated into the training dataset, it is crucial to address this challenge. Therefore, we propose selecting only a subset of LRGs as potential foreground galaxies for small Einstein radii lenses. This approach ensures that the distribution of source parameters remains consistent with the previous analysis. The variable aspect here is the selection process for the corresponding lensing galaxy associated with the mock lensed emission that have smaller Einstein radii. In Section~\ref{training_data}, we discuss the assumption that the Einstein radius ($\theta_E$) is proportional to the total integrated flux of a foreground galaxy. Here, we use the LRGs' effective radii as a proxy for their integrated flux. Under the "Applied2" setting, the training data is modified to include only LRGs with effective radii below 0.5 arcsec, paired with mock lensed emission with Einstein radii below 0.85 arcsec. However, it is important to note that the same test dataset has been utilized here as in the Vanilla and Applied1 settings. 

Fig.~\ref{fig:roc_app2} and Table~\ref{tab:roc_app2} show the results obtained from employing this strategy. We see that the most optimal performance is achieved by the ensemble of \textit{EfficientNet-B3}, \textit{EfficientNet-B4}, and \textit{DenseNet-121}, with a FP rate of $4.16 \times 10^{-4}$ when a detection threshold of 0.99 is used. This translates to the detection of 20 FPs within the 48,036 total non-lensed samples in our dataset, which is three more than the number of FPs detected using the Applied1 setting (see Figs.~\ref{fig:roc_app1} and \cref{tab:roc_app1}). Interestingly, this variation in the training dataset, by solely modifying the lens class of the training data, has influenced the number of FP detections by the model. This indicates that the challenges of strong lensing detection are complex, relying on multiple parameters. Another notable point is the 222 additional TP samples achieved using the Applied2 setting, when compared to the Applied1 and Vanilla settings. This corresponds to approximately a 0.46 per cent improvement in TPs. This presents a trade-off between selecting a model with a higher TP rate or a lower FP rate, considering that both models significantly outperform the Vanilla setting in terms of the FP rate.

\begin{table}
    \centering
        \caption{A comparison of the TP and FP rates for different CNN architectures, when the training data follows the Applied2 scenario. The best results are obtained using an ensemble of \textit{DenseNet-121}, \textit{EfficientNet-B3} and \textit{EfficientNet-B4} with a TP rate of 0.9112 and a FP rate of $4.16 \times 10^{-4}$.}
    \begin{tabular}{ccc}
    Model & TP & FP\\
    \hline
       DenseNet-121  &0.931& $4.1 \times 10^{-3}$ \\
%       \hline
       DenseNet-169 &0.932&$3.7 \times 10^{-3}$ \\
%       \hline
       EfficientNet-B3 &0.941&$3.2 \times 10^{-3}$ \\
%       \hline
       EfficientNet-B4 &0.940&$1.9 \times 10^{-3}$ \\
%       \hline
       DenseNet-121, EfficientNet-B3 &0.917&$9.3 \times 10^{-4}$ \\
%       \hline
       DenseNet-169, EfficientNet-B3 &0.918& $1.1 \times 10^{-3}$ \\
%       \hline
       DenseNet-121, EfficientNet-B4 &0.817& $6.2 \times 10^{-4}$ \\
%       \hline
       DenseNet-169, EfficientNet-B4 &0.918& $8.7 \times 10^{-4}$ \\
%       \hline
       EfficientNet-B3, EfficientNet-B4 &0.927&$8.5 \times 10^{-4}$ \\
%       \hline
       DenseNet-121, DenseNet-169 &0.916&$1.2 \times 10^{-3}$ \\
%\hline
    DenseNet-121, EfficientNet-B3, EfficientNet-‌B4&0.911 &$4.2 \times 10^{-4}$\\
%    \hline
    DenseNet-169, EfficientNet-B3, EfficientNet-B4&0.919 &$6.2 \times 10^{-4}$\\
    \hline
    \end{tabular}
    \label{tab:roc_app2}
\end{table}

\begin{figure}
\begin{center}
\begin{minipage}[b]{\columnwidth}
\centering
\includegraphics[height=2.25cm]{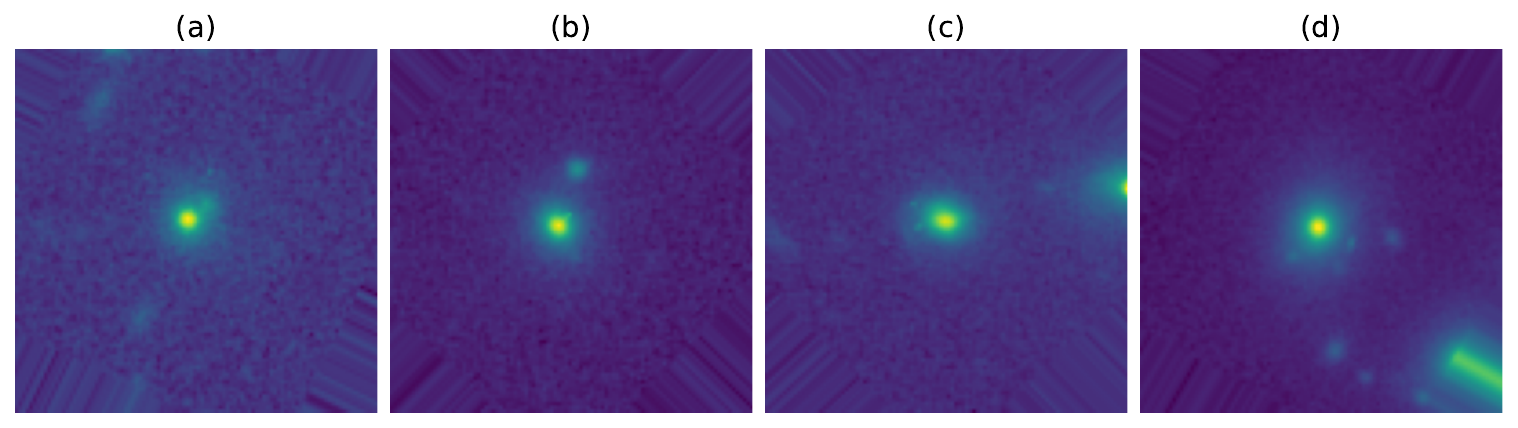}
\end{minipage}
\begin{minipage}[b]{\columnwidth}
\centering
\includegraphics[height=2.25cm]{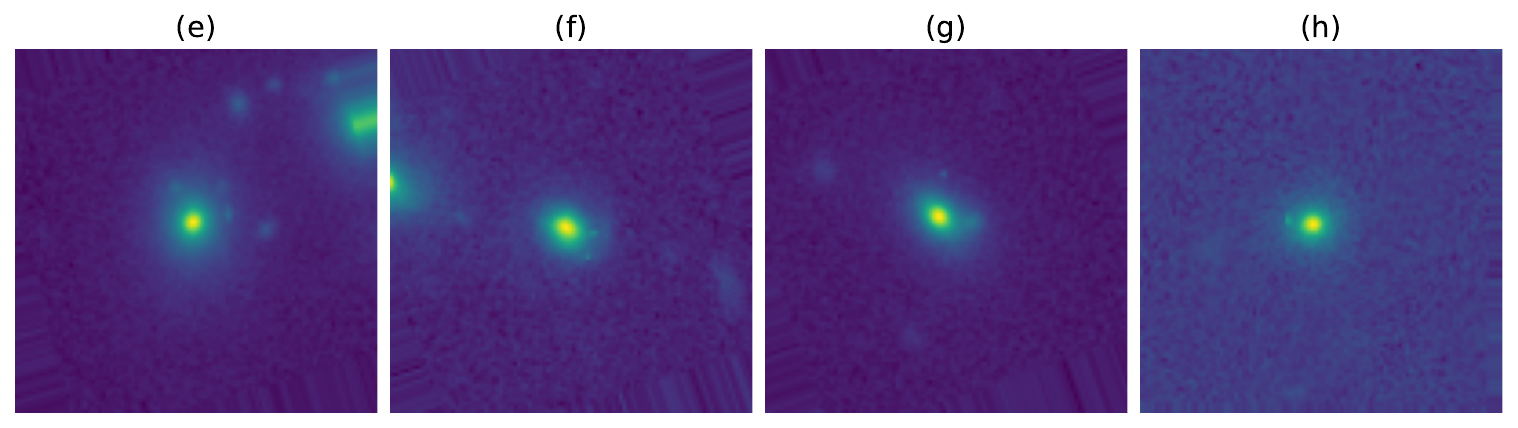}
\end{minipage}
\newpage
\caption{The 8 FP samples within the test dataset of 48,036 non-lensed samples, when considering the prediction of the Applied1 and Applied2 settings. All of the samples belong to the LRG population, which indicates that the models have learned to distinguish between spiral galaxies and lensed samples. The remaining issue is in separating the FP samples belonging to non-lensed LRGs with those that exhibit lensed emission.}
\label{fig:app1_app2_FPs}
\end{center}
\end{figure}

\subsection{Combined model} \label{discussion}
\begin{figure*}
    \centering
    \includegraphics[scale=0.7]{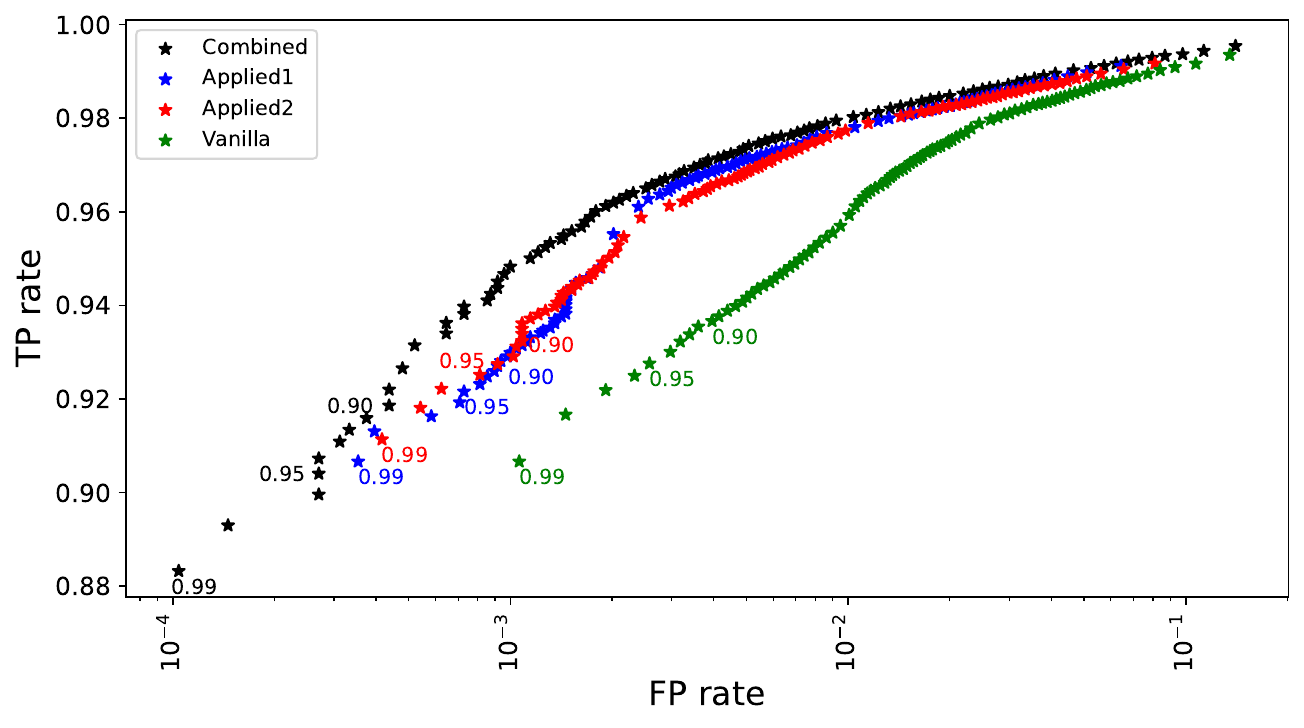}
    \caption{The ROC plot comparing the performance of each training dataset setting (Vanilla, Applied1 and Applied2). This illustrates the dependency of the TP and FP rates on the chosen threshold that separates lensed and non-lensed samples. The combined model incorporates the average lensing probability of all three training settings, demonstrating an improved FP rate, albeit with a slight reduction of the TP rate by a few per cent.}
    \label{fig:roc_combined}
\end{figure*}

Our findings have thus far highlighted the impact of modifications in the training dataset on the model's performance. We have observed that adjustments to either the lensed or non-lensed classes of data can influence the TP and FP rates, thus affecting the overall quality and reliability of the results obtained. A comparison between Tables~\ref{tab:roc_app1} and \ref{tab:roc_app2} reveals that the Applied2 setting exhibits a higher FP rate, resulting in three more FP samples compared to the Applied1 setting, within our non-lensed test dataset of 48,036 samples. Upon examination, it becomes apparent that the two models only agree on 8 FP samples, which are shown in Fig.~\ref{fig:app1_app2_FPs}. Among these, the predicted lensing probability of only 4 samples exceeds 0.99 for the Vanilla setting, indicating that if a voting strategy were employed across all models, we would identify only 4 FP samples within the entire non-lensed sample; these are labeled as (b), (c), (f), and (h) in Fig.~\ref{fig:app1_app2_FPs}. However, incorporating a strategy that averages out the predicted lensing probability for the Vanilla, Applied1 and Applied2 settings results in 5 FP samples, which now includes sample (a) in Fig.~\ref{fig:app1_app2_FPs}. This intriguing result also underscores how different ensemble approaches can impact the actual number of FPs encountered. Another interesting finding is that all of the FPs shown in Fig.~\ref{fig:app1_app2_FPs} are LRGs, which do not exhibit any spiral emission. This suggests that the primary challenge in reducing the number of FPs from the KiDS data lies not in spiral emission, but rather in distinguishing between the LRG population and contaminates that may mislead the model into interpreting them as lensed emission.

Fig.~\ref{fig:roc_combined} presents a comparison of different settings in the training dataset, namely the Vanilla, Applied1 and Applied2 settings. As previously discussed, averaging through all of these predictions yields the combined lens probability prediction, which demonstrates a better FP rate, as the models may not agree on identifying challenging samples as lensed. A more detailed view of these results is provided in Table~\ref{tab:roc_combined}, which shows the efficacy of ensemble techniques. According to our results, the combined model exhibits a FP rate of $10^{-4}$, representing an 11-fold decrease in the number of FPs compared to the Vanilla setting, a 3.5-fold improvement compared to Applied1, and a 4.1-fold improvement compared to Applied2. This significant reduction in FPs comes at the cost of a 2.3 per cent decrease in TPs when compared to the Vanilla and Applied1 settings, while compared to the Applied2 setting, the TP rate has decreased by 2.8 per cent. In the following, we provide details on how these TP and FP rates are related to the population of lensed and non-lensed samples and their underlying properties.

\begin{table}
    \centering
    \caption{A comparison of TP and FP rates for different training dataset settings (Vanilla, Applied1, and Applied2), alongside the combined model. These results highlight the superior performance of the combined model in detecting fewer FPs compared to each of the investigated training settings. The selected threshold is 0.99.}
    \begin{tabular}{ccc}
      Model   & TP & FP\\
      \hline
      Vanilla &0.906 & $1.1 \times 10^{-3}$ \\

      Applied1&0.906 & $3.5 \times 10^{-4}$ \\

      Applied2&0.911 &$4.2 \times 10^{-4}$\\

      Combined & 0.883 & $1.0 \times 10^{-4}$\\
      \hline
    \end{tabular}
    
    \label{tab:roc_combined}
\end{table}

\subsection{Parameter-space analysis} \label{parameterspace}

Utilizing the results obtained from Fig.~\ref{fig:roc_combined} and Table~\ref{tab:roc_combined}, we now investigate the parameter-space for detection, specifically examining how the Einstein radius of a lensed object or the complexity of the LRG influences the TP and FP rates. Through these experiments, our aim is to understand how different training datasets affect the types of lenses to which the trained CNN architectures are sensitive, as well as the parameters that may impact the model's ability to accurately label samples in the test dataset.

\begin{figure}
    \centering
    \includegraphics[width=\linewidth]{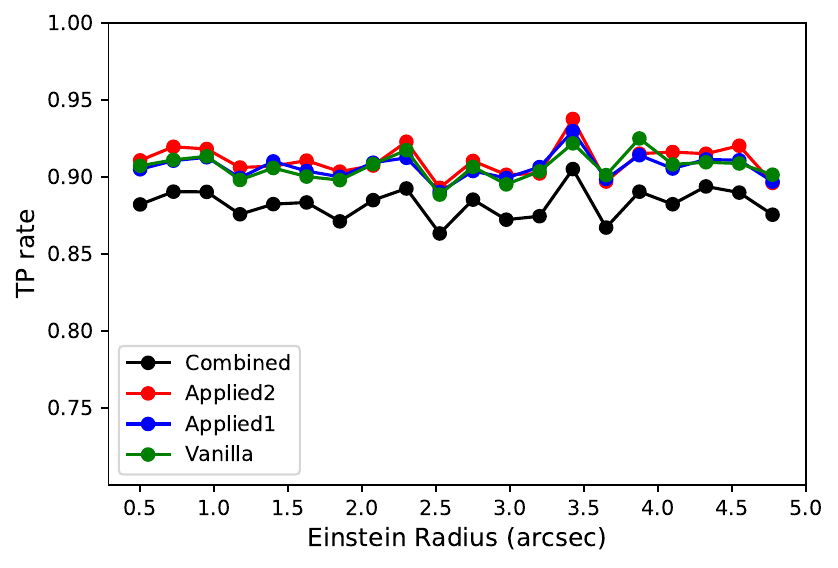}
    \caption{The distribution of TP rates as a function of the Einstein radius of the mock lenses. This reveals that the effectiveness of the detection method does not seem to be directly influenced by the size of the Einstein radius of the mock lenses. As expected from the results presented in Table~\ref{tab:roc_combined} and Fig.~\ref{fig:roc_combined}, the Combined model has the lowest TP rate, when compared to the individual models.}
    \label{fig:tp_er}
\end{figure}

Fig.~\ref{fig:tp_er} shows the TP rate as a function of the lens Einstein radius. This result indicates that although the Applied2 setting shows a slightly higher TP rate when compared to the Vanilla and Applied1 settings, this difference is consistent across all Einstein radius bins, as opposed to belonging to any specific range. On the other hand, when examining the behavior of the combined model, it becomes evident that this model exhibits a lower TP rate, when compared to the individual models. This outcome is expected because in the combined model, a sample is labeled as a lens if all three settings of Vanilla, Applied1 and Applied2 agree. This ensemble technique, achieved by averaging the predicted lens probabilities of each setting, results in a smaller set of final candidates, as is also demonstrated in Table~\ref{tab:roc_combined}.

\begin{figure}
    \centering
    \includegraphics[width=\linewidth]{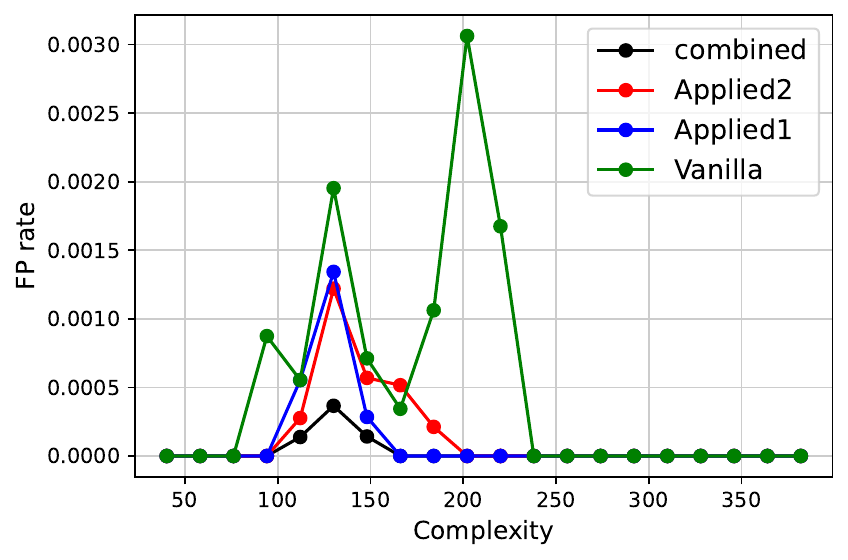}
    \caption{The FP rate as a function of the LRG compactness employed within the non-lensed population. This illustrates the influence of the distribution of the implemented training dataset on the behavior of the Vanilla, Applied1 and Applied2 settings.}
    \label{fig:fp_com}
\end{figure}

Another noteworthy comparison lies in examining the FP rate alongside the complexity of the foreground lens galaxies. As previously demonstrated, the adjustment made in the training dataset for Applied1 aims to balance the non-lensed population, ensuring that extended, complex sources are adequately represented in that class of data. A comparison between the Vanilla, Applied1 and Applied2 settings clearly highlights the impact of such adjustments on the achieved results. Fig.~\ref{fig:fp_com} illustrates a significant difference in the FP rate obtained by the Vanilla, Applied1 and Applied2 settings. An intriguing observation from Fig.~\ref{fig:fp_com} is the disparity between the FP rates of the Applied1 and Applied2 settings, despite both utilizing the exact same non-lensed samples. The sole difference between these two settings lies in the distribution of lensed samples. This observation underscores the complexity of the lens detection problem, where the behavior of the model can be influenced by numerous parameters that may initially seem unrelated.

\begin{figure*}
    \centering
    \includegraphics[width=\linewidth]{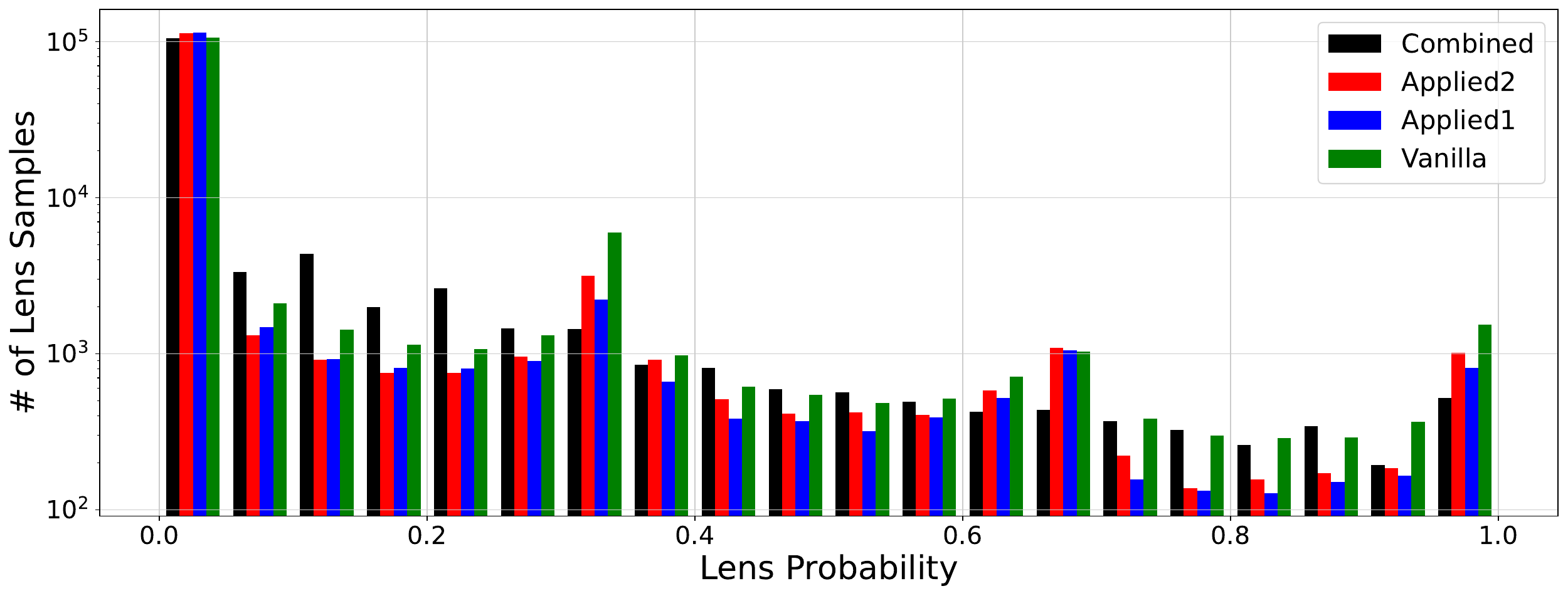}
    \caption{A comparison of the predicted lens probability for 126,000 KiDS LRG samples, as a real test dataset. As expected, the Combined model has the lowest number of predicted lens candidates, when compared to the other scenarios that have been tested in this study.}
    \label{fig:all_DR4}
\end{figure*}

\subsection{Evaluation on real KiDS data}

To evaluate the performance of our methodology on real KiDS data, we have applied the Vanilla, Applied1, Applied2 and Combined strategies to 126,000 LRGs from the KiDS DR4 \citep{RuiLi2020}. The predictions for each lensing probability bin are presented in Fig.~\ref{fig:all_DR4}. We find that the results align well with our previous analyses, such as those shown in Fig.~\ref{fig:roc_combined}. The Vanilla setting predicts the highest number of lens candidates for the probability range of [0.9--1], whereas the Applied2, Applied1, and Combined settings predict fewer samples with a high lensing probability. When using a threshold of 0.99 to identify potential lenses, the Combined strategy identifies 347 samples, whereas the Vanilla setting identifies 997 samples, indicating a threefold reduction in the number of lens candidates. This reduction is advantageous as it minimizes the need for expert visual inspection and reduces the time required for follow-up observations to confirm the strong gravitational lensing nature of these candidates. The Applied1 and Applied2 settings detect 551 and 710 samples, respectively, with a lensing probability higher than 0.99. It is crucial to assess how many genuine lensed samples are being missed among the detected candidates and whether some lenses are being overlooked in this selection process. This aspect will be addressed in our future work, where we will verify the predicted lens candidates through visual inspection.

\section{Conclusions} 
\label{conclusion}

In this study, we have investigated the complicated task of detecting gravitational lens systems through analyzing the intricate relationship between the composition of the training dataset and the performance of the detection models. Through a comprehensive analysis and experimentation, we uncovered vital insights that shed light on the multifaceted challenges and opportunities within this domain. Our findings underscored the importance of data diversity and representativeness, revealing how variations in the sample populations can cause significant influence on the behavior and efficacy of the detection models. Our study highlighted the importance of understanding the underlying reasons for model output, beyond merely assessing its performance. This iterative process often necessitates a return to the foundational step of data collection and analysis. For instance, our research revealed that the underlying distribution of LRGs in our non-lensed sample affected the number of FPs. By continuously refining the data collection methods, reassessing the dataset properties, and fine-tuning the model architectures, based on the insights gleaned from the model behaviors, we iteratively enhanced the accuracy and robustness of the detection models. 

One pivotal discovery from our research revolves around the critical need to address imbalances within the training dataset. In particular, distinguishing between extended, complex LRGs and genuine gravitational lenses posed a significant challenge. We noticed that the original test dataset, which has been widely used in the literature, has an un-balanced population of LRGs as non-lensed samples in terms of complexity. The number of complex LRGs are significantly lower in the original dataset, when compared to the compact ones. By strategically mitigating these imbalances, through techniques such as data augmentation and ensemble learning approaches, we were able to achieve notable reductions in the number of FPs, which enhanced the overall reliability of our detection model.

Beside the population of non-lensed samples in the training dataset, we also made modifications that focused on the lensed population. The adjustments changed how the lensed population is constructed, particularly concerning lenses with Einstein radii below 0.85 arcsec. We proposed to only use a subset of LRGs with a small effective radius, to act as potential foreground galaxies for these small Einstein radii lenses. The effectiveness of this modification was evaluated through experimentation, which showed an improved performance in terms of the TP and FP rates, when compared to the standard Vanilla setting.

Our examination of the FP samples obtained from the Applied1 and Applied2 settings revealed that within our test dataset of 48,036 non-lensed samples, Applied1 incorrectly labeled 17 samples as strong lenses, while Applied2 identified 20 samples as strong lenses. Further analysis showed that among those samples, Applied1 and Applied2 shared only 8 common FP samples. Interestingly, when incorporating the lensing probability predicted by the Vanilla setting into the ensemble average, only 5 FP samples remained, all of which were found to be LRGs without any evidence of spiral emission. This underscored the effectiveness of ensemble methods and highlighted the challenge of distinguishing LRG populations from potential contaminants in reducing the FP rate.

The FP rate achieved when we used the ensemble method of the Vanilla, Applied1 and Applied2 (Combined) setting was found to be $10^{-4}$, which represented an 11-fold improvement, when compared to the Vanilla setting. While this reduction in FPs came with a 2.3 per cent decrease in TPs, when compared to the Vanilla and Applied1 settings, it signified an advancement considering our primary goal of minimizing FP rates in strong gravitational lens detection algorithms.

We then evaluated our methodology using real KiDS data by applying the various strategies (Vanilla, Applied1, Applied2, and Combined) to a dataset of 126\,000 LRGs. The results showed that different strategies yielded varying numbers of high-probability lens candidates, with the Combined strategy significantly reducing the number of candidates, when compared to the Vanilla setting. This reduction is beneficial for minimizing the effort required for expert visual inspection and follow-up observations. However, it remains essential to investigate how many genuine lenses may be missed and whether some lenses are overlooked from this process. In a future work, we will focus on addressing these concerns through detailed visual inspection of the predicted lens candidates.

In conclusion, our research contributes significantly to advancing the field of gravitational lens detection by offering insights into the interplay between the training dataset, model performance, and the iterative refinement process. By emphasizing the importance of data diversity, imbalance mitigation, and continuous refinement through iterative analysis, our study provides a road-map for developing accurate and reliable detection models that are capable of unraveling the mysteries of the Universe's most intriguing phenomena.

\section*{Acknowledgements}
This work was performed using the compute resources from the Academic Leiden Interdisciplinary Cluster Environment (ALICE) provided by Leiden University.
AC was supported by the MUR PRIN2022 project 20222JBEKN with title "LaScaLa" - funded by the European Union - NextGenerationEU. This work is based on the research supported in part by the National Research Foundation of South Africa (Grant Number: 128943).

%%%%%%%%%%%%%%%%%%%%%%%%%%%%%%%%%%%%%%%%%%%%%%%%%%
\section*{DATA AVAILABILITY}
Upon reasonable request, the underlying data used for this article will be shared by the corresponding author.

%%%%%%%%%%%%%%%%%%%% REFERENCES %%%%%%%%%%%%%%%%%%

% The best way to enter references is to use BibTeX:

\bibliographystyle{mnras}
%\bibliography{bibliography} 

\bsp	% typesetting comment
\label{lastpage}
\end{document}